\documentclass[print]{mn2e}

\newcommand{\feh}{\mbox{[Fe/H]}}
\newcommand{\zh}{\mbox{[Z/H]}}
\newcommand{\afe}{\mbox{[$\alpha$/Fe]}}
\newcommand{\mgfe}{[MgFe]\arcmin}

\usepackage{graphicx}
\title[Globular Clusters in Nearby Galaxies]
{SAO-6m Telescope Spectroscopic Observations of Globular Clusters in Nearby Galaxies}
\author[M. E. Sharina, et al.]
{Margarita E. Sharina$^{1}$\thanks{E-mail: sme@sao.ru},
Rupali Chandar$^{2}$, Thomas H. Puzia$^{3}$, Paul Goudfrooij$^{4}$, 
\newauthor 
\& Emmanuel~Davoust$^{5}$\\
 $^{1}$Special Astrophysical Observatory, Russian Academy of Sciences, N.Arkhyz, KChR, 369167, Russia\\
 $^{2}$Department of Physics and Astronomy, The University of Toledo, 2801 West Bancroft Street, Toledo,
OH 43606, USA\\
 $^{3}$Herzberg Institute of Astrophysics, 5071 West Saanich Road, Victoria, BC V9E 2E7, Canada\\
 $^{4}$Space Telescope Science Institute, 3700 San Martin Drive, Baltimore, MD 21218, USA\\
 $^{5}$Laboratoire d'Astrophysique de Toulouse-Tarbes, Universit\'e de Toulouse, CNRS, 14 avenue E.~Belin, F-31400, Toulouse, France}

\begin{document}

\date{Accepted ---. Received ---}

\pagerange{\pageref{firstpage}--\pageref{lastpage}} \pubyear{2010}

\maketitle

\label{firstpage}

\begin{abstract}
We present the results of medium-resolution spectroscopy of 28 globular
clusters (GCs) in six nearby galaxies of different luminosities and morphological
types, situated in:
M33 (15 objects), M31 (3), IC10 (4), UGCA86 (4), Holmberg~IX (1), and
DDO71 (1) obtained at the Special Astrophysical Observatory 6-meter 
telescope. Measurements of Lick absorption-line indices and comparison
with SSP models enabled us to obtain their spectroscopic ages,
metallicities and $\alpha$-element to Fe abundance ratios. We found that
all old and intermediate-age GCs in our sample have low metallicities
$\zh \la -0.8$ dex. Metal-rich clusters are young and are preferentially
found in galaxies more massive than $\sim\!10^9 M_{\sun}$. The least
massive dwarfs of our sample, DDO71 and Holmberg~IX, host one
massive intermediate-age and one massive young metal-poor GC,
respectively. \afe\ abundance ratios tend to be enhanced but closer to
solar values for dwarf galaxies compared to GCs in more massive
galaxies. We analyse the age-metallicity relation for GCs in our galaxy
sample and others from the literature, and find, that 1) there is a
general trend for GCs in low surface brightness dwarf galaxies to be
more metal-poor at a given age than GCs in more massive galaxies; 2) the
GC metallicity spread is wider for more massive galaxies; 3)
intermediate-age GCs in early-type dwarf galaxies are more metal-rich
at any given age than those in irregular galaxies of similar luminosity.
\end{abstract}


\section{Introduction}

\begin{table*}
\begin{center}
\caption{Properties of our sample galaxies. The successive columns are: 
luminosity; morphological type; color excess due to Galactic extinction;
distance from the Sun; heliocentric radial velocity; distance from the
nearest massive neighbour; HI mass and total mass. All the data except
those marked by superscripts were taken from the catalogue of
Karachentsev et al. (2004). Rough total masses (marked by "::")
for DDO71 and HoIX were estimated from typical M/L ratios for dwarfs of
the corresponding morphological type.}
\label{tab:galaxies}
\scriptsize
\begin{tabular}{lcllllllr} \\ \hline\hline\noalign{\smallskip}
Galaxy         & M$_B$ & Morph.   & E(B-V) & D       & $V_h$  & D$_{\tiny MD}$ & M$_{HI}$               & M$_{tot}$                   \\
(MD)           & mag.  & Type    &mag.   & Mpc     & km/s   &  kpc     & {\tiny 10$^{9}$M$_{\sun}$}   & {\tiny 10$^{9}$M$_{\sun}$}  \\ \hline
\noalign{\smallskip}
{\bf M31}     & -21.6  & SA(s)b & 0.06   & 0.77    & -300   & --       & 5.0$^b$                      & $\sim\!340^b$               \\
	      &        &        &        &         &        &          &                              &                             \\
{\bf M33}     & -18.9  & SA(s)cd& 0.04   & 0.85    & -180   & 200      & 2.0$^c$                      & 50$^c$                      \\
(M31)         &        &        &        &         &        &          &                              &                             \\
{\bf IC~10}   & -15.6  & BCD/dIr& 0.77$^i$& 0.66   & -344   & 250      & 0.2$^d$                      & 1.6$^g$                     \\
(M31)         &        &        &        &         &        &          &                              &                             \\
{\bf UA~86}   & -17.6  & dIr    & 0.94   & 2.96$^a$&  67    & 331$^{j}$& 1.0$^e$                      & 20$^h$                      \\
(IC342)       &        &        &        &         &        &          &                              &                             \\
{\bf D71}     & -12.1  & dSph   & 0.10   & 3.5     & -129   & 210      & $\ge 0$                      & 0.1::                       \\
(M81)         &        &        &        &         &        &          &                              &                             \\
{\bf HoIX}    & -13.7  & dIr    & 0.08   & 3.7     &  46    & 70       & 0.3$^f$                      & 0.3::                       \\
(M81)         &        &        &        &         &        &          &                              &                             \\
\hline
\end{tabular}
\end{center}
{\footnotesize $^a$ The distance for UGCA86 was taken from Karachentsev et al. (2006).
Additional data: $^b$ Carignan et al. (2006), $^c$ Corbelli (2003),
$^d$ Wilcots \& Miller (1998), $^e$ Rots (1979), $^f$ Yun et al. (1994),
$^g$ Mateo (1998), $^h$ Stil et al. (2005), $^i$ Massey \& Armandroff (1995) , $^j$ this paper.}
\end{table*}

Star clusters (SCs) are fundamental building blocks of galaxies (Lada \&
Lada, 2003). Stellar groups and associations, open clusters (OCs), globular 
clusters (GCs), and super star clusters (SSCs) are members of one family 
(e.g. Elmegreen 2002, Kroupa \& Boily, 2002). Their main differences reside in the
density and pressure of the progenitor molecular clouds and their
environmental conditions. There is no strict difference between OCs
and GCs in our Galaxy. Their ranges in age, metallicity, mass and radius
intersect. In general, OCs are younger and more metal-rich
than GCs, and reside in the disc (Harris 1996, Dias et al. 2002). SSCs
are young populous clusters and probable progenitors of compact GCs
exceeding $\sim 10^5 M_{\sun}$ within a radius of 1--2 pc. There are
nuclear clusters and SSCs found in undisturbed late-type and in
interacting starburst galaxies and regions of galaxies with signatures
of large-scale shock compression of the interstellar medium (e.g. Arp \&
Sandage 1985, Figer et al. 1999, Crowther et al. 2006). The formation
of massive gravitationally bound star clusters in dwarf galaxies is a
natural consequence of the high mass-to-luminosity ratios (M/L) and
hence high virial densities (from stars, gas and dark matter) and
ambient pressures (Elmegreen \& Efremov 1997, Ashman et al. 1994).

According to the cold dark matter cosmological paradigm globular
clusters formed from 3~$\sigma$ density fluctuations in
low-mass ($\sim\!10^6\! -\!10^8$) dark matter halos, before merging into
larger structures (Peebles 1984, Mashchenko \& Sills 2005a,b; Moore et
al. 2006). The hosts of these halos were probable progenitors of the
present-day dwarf galaxies. All their representatives in the Local Group
and beyond, resolved into individual stars up to now, contain old
stellar populations with the only probable exception of tidal dwarfs
(see e.g. Grebel 1999). Ultra-faint dwarf galaxies and the most extended
GCs have similar densities, luminosities, and sizes. However, the former
are strongly dark matter dominated, with again the exception of tidal
dwarfs (Barnes \& Hernquist, 1992). A lower limit to the halo mass of a
small galaxy with GCs is hard to determine observationally. For this one
should have good statistics of GCs in the lowest-mass isolated galaxies.

Since the chemical composition of stars in the Galaxy and its satellites
are very different (Venn et al. 2004, Pritzl et al. 2005), the scenario
of pure hierarchical merging of small fragments does not explain their
formation process. Present day dwarf and giant galaxies seem to have
experienced very different chemical evolutions, and, additionally, the
percentage of late consecutive merging events was small. Dwarf
satellites were probably captured by our Galaxy without significant
bursts of star formation (SF), as in the Sagittarius spheroidal (dSph,
Ibata et al. 1994).
Outside the Local Group (LG) Hubble Space Telescope (HST) observations
revealed signatures of numerous relatively recent galaxy
merging and accretion events, and a plethora of young massive SCs
originated in mergers of gas-rich hosts (e.g. Whitmore et al. 1999,
Harris 2001 and references therein). The formation of SCs is thus
intimately linked to their parent galaxy's evolution. A good method to
investigate the assembly history of galaxies is chemical tagging of
stars and representatives of the brightest simple stellar populations,
e.g. GCs, in galaxies of different morphological types, masses, and
luminosities (West et al., 2004, and references therein). A suitable
laboratory for testing cosmological theories is the close neighbourhood
of the Galaxy, where clusters can be observed in detail.

In this work we analyse spectra of 28 GCs in six galaxies of different
luminosities situated within $ \sim4$ Mpc
in different group environments:
1) the giant spiral neighbour of our Galaxy M~31;
2) the intermediate-luminosity spiral M~33;
3) IC~10, a starburst dwarf irregular (dIrr) member of the LG;
4) UGCA86, a Magellanic-type gas-rich dwarf satellite of IC342, with a complex
structure including two bright starburst regions in the visible and a rotating disc as well as a spur in H{\sc i}
(Stil et al. 2005 and references therein); and two low surface brightness (LSB) dwarf companions of M81,
5) the spheroidal DDO~71, and 
6) the tidal dIrr Holmberg~IX (Ho~IX) (van den Bergh 1959). 
Luminosities, heliocentric radial velocities and distances for the
galaxies of our study are listed in Table~\ref{tab:galaxies}. The
absolute magnitudes of UGCA86 and IC~10 are uncertain due to a high
Galactic extinction and an unknown internal contribution. The distances
to M~31, M~33, and IC~10 were derived from the luminosity of Cepheids
(Karachentsev et al. 2004, and references therein). The projected
positions of UGCA86, and DDO~71 with respect to the Galaxy are known
from the visual luminosity of their stars on the tip of the red giant
branch (RGB). The distance to HoIX is uncertain. It is taken equal to
the distance to M81, which was first derived by Georgiev et al. (1991a)
from the visual magnitude of the brightest blue and red supergiant stars
on SAO 6m-telescope photographic plates. It was not possible to improve
the distance using high-resolution HST images because of the absence of
clear signs of RGB in HoIX (see Karachentsev et al., 2002). Red giants
are randomly distributed within the boundaries of the galaxy and may
belong to M81 (Makarova et al. 2002, Sabbi et al. 2008). HoIX is tightly
bound to M81 and, additionally, is in active tidal interaction with its
large neighbour, as evidenced by the HI distribution pattern in the
centre of the M81 group (Yun et al., 1994).

The paper is organized as follows. In Section~2 we describe the
selection of GC candidates for the spectroscopic survey. In Section~3
the methods of observation and spectra reduction are explained. The
evolutionary parameters obtained from the measured absorption-line Lick
indices are listed and discussed in Section~4. An interpretation of the
data is given in Section~5. We formulate our conclusions in Section~6.

\section{Selection of candidates for the spectroscopic survey}
\subsection{Search for and Photometry of Star Clusters in IC~10 and UGC~A86}

\begin{figure*}
\includegraphics[width=0.9\textwidth]{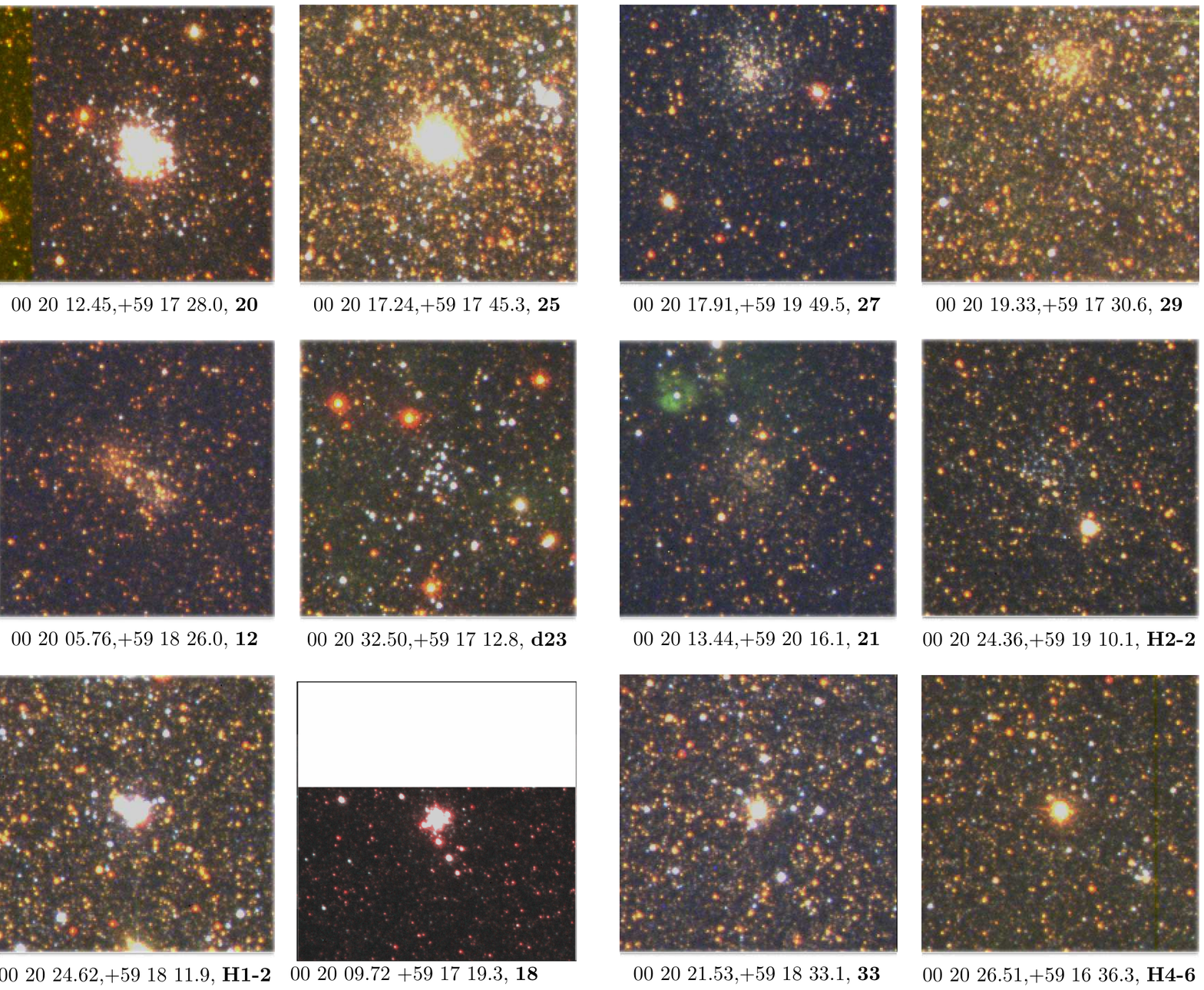}
\vspace{-9cm}
\caption{Examples of the HST/ACS images for the star clusters found in IC10:
bright compact globular morphology (20 and 25);
relatively bright intermediate morphology (27 and 29);
open morphology (double 12, bright d23, faint red  21, and faint blue  H2-2
from the list of Hunter 2001); very compact (H1-2 and 18); star-like (33 and H4-6).
The size of the images is $20\arcsec\times 20\arcsec$.}
\label{examples}
\end{figure*}
\begin{table}
\caption{HST images used for the search and photometry of star cluster candidates.}
\label{tab:hst}
\scriptsize
\begin{tabular}{llllr} \\ \hline\hline\noalign{\smallskip}
Prop ID & PI Name & Detector & Filters      & Exposure Time \\ \hline
\noalign{\smallskip}
\noalign{\bf IC10}
\noalign{\smallskip}
9683   &  Bauer   & ACS/WFC  & F555W & 16x1240              \\
       &          &          & F814W  & 8x1190               \\
10242  &  Cole    & ACS/WFC  & F606W  &  1080, 1080          \\
       &          &          & F814W  &  1080, 1080          \\
       &          &          & F435W  &  1020, 1020          \\
\noalign{\smallskip}
\noalign{\bf UGCA86}
\noalign{\smallskip}
9771   & Karachentsev& ACS/WFC  & F606W  & 1200      \\
       &             &          & F814W  &  900      \\
\hline
\end{tabular}
\end{table}
\begin{table}
\caption{Literature sources for coordinates and magnitudes of the studied GCs.
The acronym is given in the first column.}
\label{tab:lit}
\scriptsize
\begin{tabular}{ll} \\ \hline\hline\noalign{\smallskip}
Abbreviation         & Reference  \\ \hline
\noalign{\smallskip}
\noalign{\bf M31}
\noalign{\smallskip}
CCS85  & Crampton et al. (1985) \\
MKKSS98 & Mochejska et al. (1998) \\
Bol   & Battistini et al. (1980, 1987) \\
BHB2000 & Barmby et al. (2000) \\
\noalign{\smallskip}
\noalign{\bf M33}
\noalign{\smallskip}
CS82  & Christian \& Schommer (1982, 1988) \\
KM60  & Kron \& Mayall (1960)         \\
MKKSS98 & Mochejska et al. (1998) \\
MD78  & Melnick \& D'Odorico (1978) \\
KK97  & Kunchev \& Kaltcheva (1997)  \\
CBF   & Chandar, Bianchi, \& Ford (1999) \\
\noalign{\smallskip}
\noalign{\bf IC10}
\noalign{\smallskip}
H X-X & Hunter (2001) \\
\noalign{\smallskip}
\noalign{\bf DDO~71 and HoIX}
\noalign{\smallskip}
SPM2005 & Sharina, Puzia, \& Makarov (2005) \\
\hline
\end{tabular}
\end{table}

We searched for SC candidates in UGC~A86 and IC~10 by eye on
high-resolution HST images (see Table~\ref{tab:hst}) without taking into
account literature sources. We used data from the HST archives for two
different programs with short and long exposure times. Additionally, two
GC candidates far from the centre of IC~10 were indicated by
N.A.~Tikhonov (private communication), one of which (\#36), 
observed by us spectroscopically, was found by Tikhonov on SAO
6m-telescope photographic plates. The coordinates of the objects
were measured using the Hubble Legacy Archive interactive display
service.

In spite of the enormous Galactic and inhomogeneous internal extinctions,
saturating faint stars, the structure of the galaxies is seen in great detail.
Only the central $\sim\!3\!-\!5$\arcmin\ of active SF in IC10 is covered by 
the HST images.
The images overlap, and we managed to find additional SCs on the deeper images.
Star clusters from globular to open morphology were found just from their appearance,
without paying attention to their colors. Then their images in different filters were
examined, cleaned simultaneously
of foreground stars and background galaxies, based on the cluster's mean color,
and measured in a standard manner (see e.g. Sharina et al. 2008), after approximating and
subtracting the Galactic diffuse background around each object. The centre 
was chosen by eye, and then re-determined using the {\small MIDAS} 
program FIT/ELL3 in the context ``SURFPHOT''. Photometry was made in circular apertures with radii growing
from 1 pixel up to the limiting radius of the object (growth curve limit).
The background was estimated in a circular area around the cluster. 
The photometric errors in each circular aperture were calculated
taking into account its total flux (I) in ADU, and area in arcsecond$^2$: $dI = \sqrt{I/g + S \cdot RON^2}$,
where $g$ is gain in e$^-$/ADU, S is area in arcsecond$^2$, 
and RON is read-out-noise in e$^-$/arcsecond$^2$.
To transform the surface photometry results into the standard Johnson-Cousins
system we use the zeropoints and calibration coefficients from Sirianni et al. (2005).

In UGCA86 we only searched for compact spherical clusters (see Sharina et al. 2005 for the criteria).
Many SCs of open morphology are also seen on the ACS images of UGCA86.
So, the work may be continued by an inventory and the photometry of these objects.

A separate paper will be devoted to a detailed discussion of the
photometric and structural properties of the SCs, as well as the
issues of open -- globular cluster dichotomy. Here we list their basic
parameters, which provide information about their nature. The list of
SCs with the corresponding equatorial coordinates, V magnitude, colors,
and central surface brightness is given in Tables~\ref{ic10} and
\ref{ua86} of the Appendix. The true-color HST images of the clusters in
IC10 and the two-color ones for UGCA86 may be obtained from the
SAO~RAS ftp-site\footnote{ftp://ftp.sao.ru/pub/sme/6m-spectr/Images/}.

HST ACS/WFC images for 12 of 69 clusters found in IC10 are shown in
Figure~\ref{examples}, which illustrates the diversity of morphologies,
shapes, colors, stellar content of the cluster population. A brief
inspection of our photometric data shows that the number of bright
compact SCs in UGCA86 is larger than in IC10. Almost all objects found
in UGCA86 have globular morphology. The compact GCs in IC10 are (see
Tab.~\ref{ic10}): 4, 16, 18, d9, 19, 20, 24, 25, 30, H2-2, H1-2.
\subsection{M31, M33, DDO71, and HoIX}

Candidates for spectroscopic observations in M31, M33, DDO71, and HoIX
were taken from the literature sources listed in Table~\ref{tab:lit},
where the acronyms were taken from
SIMBAD\footnote{http://simbad.u-strasbg.fr/simbad/}. The coordinates and
magnitudes of the observed GCs are summarised in Table~\ref{tab:coord}.

\begin{table}
\centering
\caption{Coordinates and magnitudes of the observed star clusters.
Ordinal numbers of star clusters in M33, marked in bold font in the first column of this table,
were used in Tables \ref{tab:evpar}, \ref{lickind1} and  \ref{lickind2}, and on Fig.~\ref{DiagD}, \ref{DiagD2}.}
\label{tab:coord}
\scriptsize
\begin{tabular}{llll} \\ \hline\hline\noalign{\smallskip}
Object          & R.A. (2000) Dec.       & V   & I, or B   \\
		&                        &     &     \\
\hline
\noalign{\smallskip}
\noalign{\bf IC10}
\noalign{\smallskip}
18       & 00 20 09.66 +59 17 19.1& 18.48& 16.78$^I$               \\
20       & 00 20 12.44 +59 17 27.9& 17.80& 16.04$^I$            \\
25       & 00 20 17.24 +59 17 45.3& 17.80& 16.32$^I$             \\
36       & 00 20 27.5  +59 13 25.9& .... & ....                           \\
\noalign{\smallskip}
\noalign{\bf DDO71=KDG63}
\noalign{\smallskip}
KDG63-3-1168  & 10 05 07.2 +66 33 30.0 & 20.95& 19.84$^I$                         \\
\noalign{\smallskip}
\noalign{\bf HoIX}
\noalign{\smallskip}
HoIX-4-1038   & 09 57 40.0 +69 03 25.0 & 19.55& 19.00$^I$                         \\
		  \noalign{\smallskip}
\noalign{\bf UGCA86}
\noalign{\smallskip}
13       & 03 59 48.2 +67 08 19.0 & 23.00& 21.50$^I$               \\
20       & 03 59 49.9 +67 06 49.2 & 22.20& 21.12$^I$                \\
22       & 03 59 50.3 +67 08 36.8 & 23.32& 21.71$^I$               \\
32       & 03 59 56.5 +67 06 11.7 & 19.28& 17.31$^I$               \\
\noalign{\smallskip}
\noalign{\bf  M33}
\noalign{\smallskip}
{\bf 2},MKKSS27         & 01 34 00.3 +30 37 47.1& 16.05 & 16.66$^B$   \\ 
{\bf 3},CBF99           & 01 34 00.5 +30 41 21.7& 18.154& ....               \\
{\bf 4},KK6             & 01 34 01.0 +30 39 38.1& 16.30 & 17.03$^B$          \\
{\bf 5},CBF120          & 01 34 01.3 +30 39 23.1& 18.169& ....               \\
{\bf 6},CSR14, CBF98    & 01 34 02.5 +30 40 39.3& 16.48 & 17.46$^B$          \\
{\bf 7},MKKSS33         & 01 34 02.9 +30 43 20.1& 16.31 & 17.02$^B$          \\
{\bf 8},CBF119          & 01 34 06.3 +30 37 30.1& 18.247& ....                 \\
{\bf 9},CS R12,CBF116   & 01 34 08.0 +30 38 38.2& 16.38 &  17.41$^B$           \\ 
{\bf 10},CS U89,CBF56   & 01 34 14.0 +30 39 29.6& 18.41 & ....                 \\
{\bf 11},CS U78,MKKSS42 & 01 34 11.4 +30 41 27.7& 18.03 & ....                 \\
{\bf 12},CS U82,MKKSS44 & 01 34 14.2 +30 39 58.2& 19.2  & 19.3$^B$             \\
{\bf 13},CS U83,MKKSS41 & 01 34 10.9 +30 40 30.0& 18.5  & 19.3$^B$               \\
\noalign{\smallskip}
{\bf 1},CBF129          & 01 33 56.1 +30 38 40.2& 17.38 & ....                \\
{\bf 14},CBF118         & 01 34 06.3 +30 37 25.7& 17.945& ....           \\
{\bf 15},CS~U73, CBF~152 & 01 34 08.7 +30 42 55.1& 18.55& 18.87$^B$      \\
\noalign{\smallskip}
\noalign{\bf M31}
\noalign{\smallskip}
MKKSS61, CCS74 & 00 45 07.2 +41 40 32.2& 18.12& ....      \\
MKKSS58    & 00 45 03.3 +41 40 05.6& 18.63& ....               \\
MKKSS72    & 00 45 13.8 +41 42 26.1& 18.40&                      \\
\hline
\end{tabular}
\end{table}
\section{Spectroscopic observations and data reduction}

\begin{table}
\caption{Journal of observations. LS refers to the long-slit mode of spectroscopic observations, and MS refers to the multislit one.}
\label{tab:obslog}
\scriptsize
\begin{tabular}{lcllc} \\ \hline\hline\noalign{\smallskip}
Object         & Date            & Exposure (s)  & Seeing (\arcsec) \\ \hline
\noalign{\smallskip}
IC10(MS)        & 15/09/04      & 5 x 1200      &  1.5 \\
IC10 GC1 (LS)   & 11/09/07      & 5 x 1200      &  1.7 \\
\noalign{\smallskip}
DDO71 (LS)      & 17/01/07      & 10 x 900      &  1.7  \\
HoIX  (LS)      & 17/01/07      & 3 x 900       &  1.7  \\
UGCA86 (MS)     & 12/09/07      & 1200, 700     &  2.0  \\
\noalign{\smallskip}
M33 (MS)& 21/08/06     & 3 x 900        & 2.0    \\
M33 (MS)& 10/09/07     & 5 x 1200       & 2.6    \\
M33 (MS)& 11/09/07     & 4 x 1200       & 3.0    \\
\noalign{\smallskip}
M31 (MS)& 10/09/07     & 7 x 1200       & 3.0    \\
\noalign{\smallskip}
\noalign{\bf Lick standard stars}
\noalign{\smallskip}
HD132142         & 15/09/04     &  10            & 1.5   \\
HD4744           & 15/09/04     &  30            & 1.5   \\
HR1015           & 15/09/04     &  10            & 1.0   \\
HD67767          & 16/12/04     &  120           & 3.0   \\
HD72184          & 16/12/04     &  20 x 2, 40    & 3.0   \\
HD74377          & 15/12/04     & 120, 240       & 3.0   \\
HR0964           & 16/12/04     & 20, 40         & 3.0    \\
HR3422           & 15/12/04     & 120 x 2        & 3.0    \\
HR3427           & 15/12/04     & 120 x 2        & 3.0    \\
HR3428           & 15/12/04     & 120 x 2        & 3.0    \\
HD4744           & 10,11,12/09/07  & 20          & 1.7 -- 2.6 \\
		 &  21/08/06       & 10          & 2.0         \\
HD2665           & 10,11,12/09/07  & 20          & 1.7 -- 2.6 \\
		 & 17/01/07        & 10          & 2.0          \\
HD7010           & 10,11,12/09/07& 20            & 1.7 -- 2.6 \\
HR4435           & 17/01/07        & 6,6         & 2.0          \\
HD132142         & 17/01/07        & 10,10       & 2.0          \\
\noalign{\smallskip}
\noalign{\bf Spectroscopic standard stars}
\noalign{\smallskip}
BD+25d4655       & 10,11,12/09/07 & 60          & 3.0     \\
Feige 34         & 17/01/07        & 10          & 2.0     \\
		 & 15,16/12/04     & 120, 240    & 3.0     \\
HZ4              & 15/09/04        & 60 x 2      & 1.5     \\
\hline
\end{tabular}
\end{table}

The spectroscopic data were obtained with the SCORPIO spectrograph
(Afanasiev \& Moiseev 2005; see also Afanasiev et al. 2005, and the
instrumental
web-page\footnote{http://www.sao.ru/hq/lsfvo/devices/scorpio/scorpio.htm
l} for detailed information about the spectrograph), installed at the
prime focus of the SAO-6m telescope of the Russian Academy of Sciences
in two modes: with longslit (LS) and multislit (MS) units (see the
journal of observations in Table~\ref{tab:obslog} for details). In the
MS mode, SCORPIO has 16 movable slits (1\farcs2 x 18\farcs0) in the
field of 2\farcm9 x 5\farcm9 in the focal plane of the telescope. The
slit width was 6\arcmin x1\arcsec\ in the long slit mode. We used the
CCD detector EEV42-40, the grism VPHG1200g (1200 lines/mm) with a
spectral resolution $\sim\!5$ \AA. The spectral range is between 3800
and 6000\AA. It changes slightly with the Y position of the object
within the field of view in the MS mode. A major difficulty of our
observations with the MS device was to set the slits correctly for the
following reasons. First, the area free of aberrations around the centre
of the camera is only $\sim 3\times3$~deg. Targets must fall accurately
in the centre of the slit, otherwise radial velocities are rough for
star-like objects due to the inhomogeneous illumination of the slits.
Second, there are restrictions on the possible angles of rotation of the
platform mounted at the prime focus (30 degrees are unreachable). Third,
the slits should be oriented as close as possible to the direction of
the atmospheric dispersion line. The settings of the slits for the
objects observed using the MS mode may be obtained from the SAO~RAS
ftp-site\footnote{ftp://ftp.sao.ru/pub/sme/6m-spectr/}.

Before applying the standard procedure of primary reduction to each
two-dimensional spectrum obtained in the MS mode, one needs to correct
the geometric field distortions as described in Sharina et al. (2006b).
The software was written in {\small IDL} by V.L.~Afanasiev.

\begin{figure*}
\vspace{-1cm}
\begin{tabular}{p{0.50\textwidth}p{0.50\textwidth}}
\hspace{-1cm}
\includegraphics[width=0.73\textwidth]{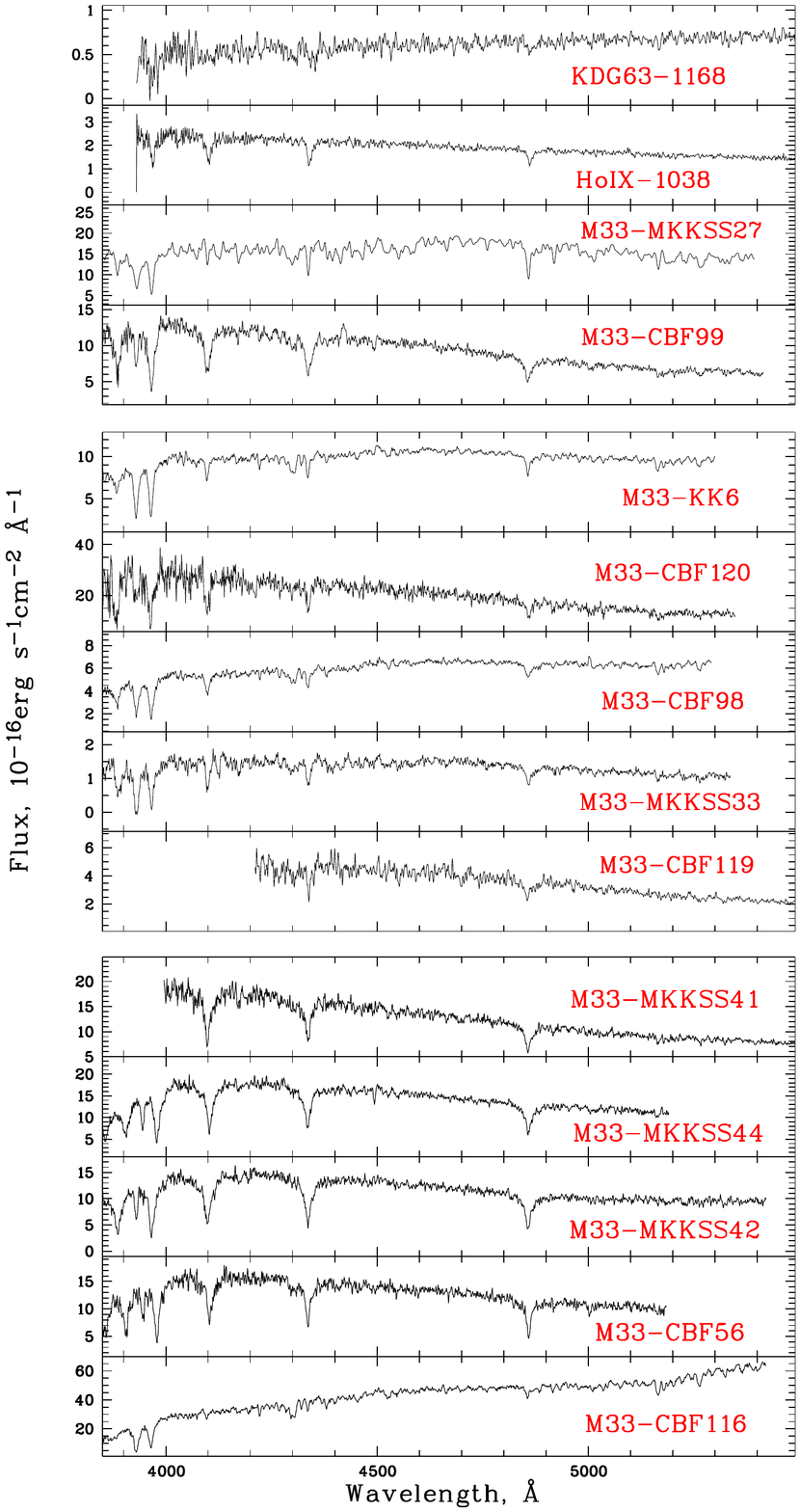}&
\hspace{-1cm}
\includegraphics[width=0.73\textwidth]{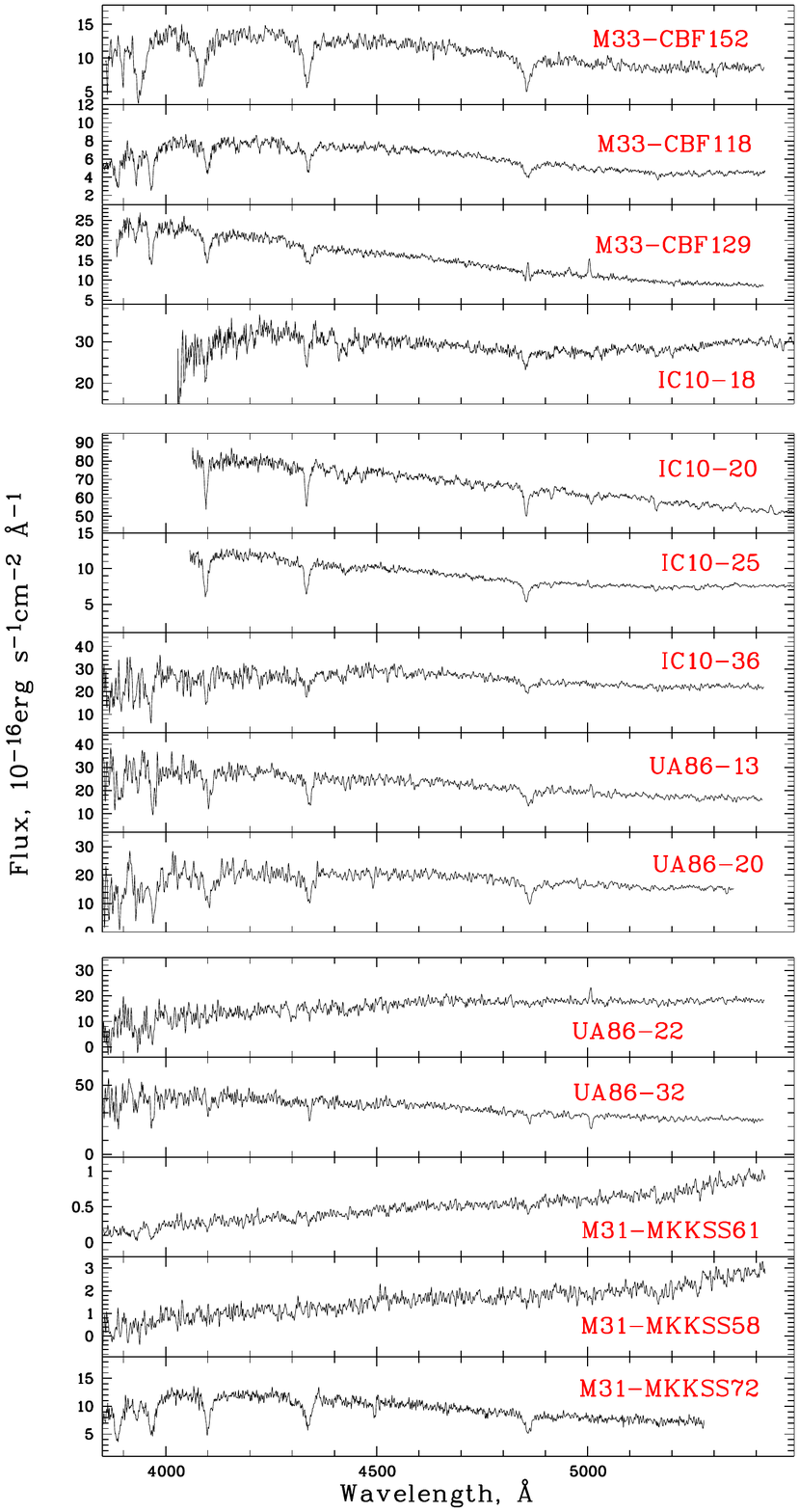}\\
\end{tabular}
\caption{Spectra of our sample GCs for which evolutionary parameters 
were obtained (see Tab.~\ref{tab:evpar}).}
\label{spectra}
\end{figure*}

The standard data reduction and analysis of the long-slit observations
and the analysis of the MS observations were performed using the
European Southern Observatory Munich Image Data Analysis System (MIDAS)
(Banse et al., 1983), and the Image Reduction and Analysis Facility
(IRAF) software system\footnote{http://iraf.noao.edu/}. The dispersion
solution determines the accuracy of the wavelength calibration which is
$\sim$0.08 \AA. A typical dispersion was 0.88 \AA/pix. The wavelength
zeropoint shifted during the night by up to 2 pixels. It was checked
using the HgI~$\lambda4358$, and [OI]~$\lambda5577$ strong night-sky
lines in the dispersion-corrected spectra. Optimal extraction of the
spectra (Horne, 1986) was made using the IRAF procedure {\it apsum}.
After wavelength calibration and sky subtraction, the spectra were
corrected for extinction and flux-calibrated using the observed
spectrophotometric standard stars (Oke, 1990). Finally, all
one-dimensional spectra of each object were summed to increase the S/N
ratio. The resulting spectra are shown in Figure~\ref{spectra}.

\begin{table*}
\centering
\caption{Correction terms of the transformation to the Lick/IDS
standard system: $ I_{\rm Lick}=I_{\rm measured}+c $ for different observational dates and instrument configurations :
10,11,12/09/07 (multislit); 17/01/07 (longslit).}
\label{tab:zeropoints}
\begin{tabular}{lrrrrcr} \\ \hline
Index       & c, err (10/09/07)&  c, err (11/09/07)& c, err (12/09/07)&   c, err (17/01/07)&Index range& units \\ \hline
\noalign{\smallskip}
CN1         & $-$0.008, 0.012  &  $-$0.019, 0.007  & $-$0.012, 0.011  &   $-$0.028, 0.012  &[-0.07 -- 0.4]   & mag    \\
CN2         & $-$0.007, 0.011  &  $-$0.015, 0.003  & $-$0.012, 0.010  &   $-$0.022, 0.005  &[-0.05 --0.4]    & mag    \\
Ca4227      & $-$0.274, 0.316  &  $-$0.260, 0.200  & $-$0.205, 0.207  &      0.069, 0.228  & [0.17 -- 2.5]    &  \AA   \\
G4300       & $-$0.775, 0.300  &  $-$0.958, 0.214  & $-$0.913, 0.294  &      0.033, 0.640  & [5.2 -- 7.0]     &  \AA   \\
Fe4384      &    0.048, 0.200  &  $-$0.271, 0.205  & $-$0.252, 0.230  &      0.391, 0.309  & [2.3 -- 4.0]     &  \AA   \\
Ca4455      &    0.197, 0.230  &     0.480, 0.240  &    0.286, 0.205  &      0.327, 0.523  & [1.0 -- 9.1]     &  \AA   \\
Fe4531      &    0.336, 0.340  &  $-$0.115, 0.450  &    0.467, 0.210  &      0.199, 0.163  & [1.0 -- 5.2]     &  \AA   \\
Fe4668      &    0.082, 0.200  &     0.090, 0.200  &    0.082, 0.200  &      0.242, 0.314  & [0.2 -- 10.4]    &  \AA   \\
H$\beta$    & $-$0.433, 0.110  &  $-$0.187, 0.211  & $-$0.303, 0.198  &   $-$0.308, 0.268  & [0.8 -- 1.1]     &  \AA   \\
Fe5015      &    0.280, 0.210  &     0.617, 0.312  &    0.685, 0.303  &      0.524, 0.010  & [1.6 -- 7.3]     &  \AA   \\
Mg$_1$      & $-$0.046, 0.003  &  $-$0.009, 0.020  & $-$0.009, 0.014  &      0.001, 0.018  & [-0.01 -- 0.21]   &  mag    \\
Mg$_2$      &    0.034, 0.006  &     0.033, 0.021  & $-$0.002, 0.017  &   $-$0.001, 0.021  & [0.01 -- 0.36]   &  mag    \\
Mgb         &    0.221, 0.126  &     0.262, 0.336  &    0.279, 0.250  &   $-$0.098, 0.082  & [0.3 -- 3.9]     & \AA  \\
Fe5270      &    0.427, 0.128  &     0.654, 0.400  &    0.511, 0.101  &      0.175, 0.199  & [0.7 -- 4.3]     & \AA  \\
Fe5335      & $-$0.238, 0.188  &     1.051, 0.250  & $-$0.134, 0.313  &   $-$0.237, 0.305  & [0.1 -- 3.9]     & \AA  \\
Fe5406      &    1.121, 0.301  &     1.002, 0.200  &    1.254, 0.470  &      0.037, 0.305  & [0.2 -- 2.9]     & \AA  \\
H$\delta_A$ & $-$0.212, 0.310  &  $-$0.362, 0.301  & $-$0.269, 0.430  &   $-$0.122, 0.438  & [-7.6 -- 0.6]     & \AA  \\
H$\gamma_A$ & $-$2.932, 0.202  &  $-$2.964, 0.299  & $-$2.930, 0.402  &   $-$0.199, 0.432  & [-11.6 -- -3.6]     & \AA  \\
H$\delta_F$ & $-$0.334, 0.220  &  $-$0.355, 0.305  & $-$0.243, 0.198  &      0.123, 0.280  & [-2.1 -- 1.1]     &  \AA  \\
H$\gamma_F$ & $-$0.273, 0.213  &  $-$0.135, 0.199  & $-$0.261, 0.233  &   $-$0.325, 0.093  & [-3.5 -- -0.7]     &  \AA  \\
\noalign{\smallskip}
\hline
\end{tabular}
\end{table*}
We calibrated the instrumental absorption-line strengths measured in
each mode onto the Lick standard system (Worthey 1994; Worthey \&
Ottaviani 1997) by observing Lick standard stars from the list of
Worthey et al. (1994). The Lick zeropoints were calibrated each night to
minimise systematic errors in the measurements of absorption-line
indices (Table~\ref{tab:zeropoints}). The radial velocities of the
objects were determined using the method of Tonry \& Davis (1979) using
the observed Lick standard stars. The derived heliocentric radial
velocities are listed in Table~\ref{tab:evpar}. The Lick indices
measured in the spectra of GCs and brought into correspondence with the
Lick system are listed in the Appendix in Tables~\ref{lickind1} and
\ref{lickind2}.

\begin{table*}
\centering
\caption{Heliocentric radial velocities and estimated evolutionary 
parameters of our sample star clusters. The columns contain the
following data: (1) Identifier of each cluster; (2) approximate S/N per
pixel measured at 5000\AA\ of the initial one-dimensional spectrum, not
degraded to the resolution of the Lick system; (3) heliocentric radial
velocity; (4),(5),(6) age, \afe\, and \zh\ estimated with TMB03 models;
(7), (8) age and metallicity, $Z$, in units of solar metallicity derived
with GD05 models; (9), (10) age and \feh\ calculated using the method of
full spectrum fitting and the Vazdekis (1999) models. A colon is
used when the errors of \feh , age, and \afe\ are larger than 0.4 dex, 4
Gyr, 0.3 dex, respectively. A check mark "v" indicates in all the cases,
except two, the results obtained with TMB03 models, and consistent with
those estimated with the other approaches. For HoIX-4-1038 and
UGCA86-13, the GD05 and the Vazdekis (1999) models give compatible
results (see text for details).}
\label{tab:evpar}
\scriptsize
\begin{tabular}{lcc|cccc|ccc|cc} \\ \hline\hline\noalign{\smallskip}
Object  & S/N   & V$_h$                       & &     age     &     \afe     &       \zh    & &  age$^G$     &   Z$^G$     &  age$^V$      & \feh$^V$       \\
	&        & km/s                       & &     (Gyr)   &     (dex)    &       (dex)  & & (Gyr)       &(Z$_{\sun}$)  & (Gyr)         & (dex)          \\
(1)     & (2)   &  (3)                        & &   (4)       &   (5)        &   (6)        & &  (7)         &  (8)        &   (9)         &  (10)          \\
\hline
\noalign{\smallskip}
{\bf IC10}                      &             & &             &              &              & &             &              &               &               \\
18 & 35                         & -452$\pm$30 &v& 12$\pm$3    & 0:           & -1.4$\pm$0.2 & & 8.7$\pm$0.8 & 0.05$\pm$0.10&   12:         & -1.7:         \\
20 & 42                         & -419$\pm$30 &v&  5$\pm$1    & 0.1$\pm$0.25 & -1.1$\pm$0.1 & & 6.2$\pm$0.2 & 0.05$\pm$0.05&   9.7$\pm$0.3 & -1.60$\pm$0.04\\
25 & 75                         & -340$\pm$30 &v&0.7$\pm$0.1  & 0.03$\pm$0.16& -0.2$\pm$0.1 & & 2.0$\pm$0.9 & 0.28$\pm$0.28&   0.1:        & -0.10$\pm$0.03\\ 
36 & 20                         & -350$\pm$30 &v&  8$\pm$3    & 0.4:         & -1.3$\pm$0.3 & & 5.9$\pm$1.2 & 0.05$\pm$0.18&   4.5$\pm$0.3 & -1.7:         \\
				&             & &             &              &              & &             &              &               &               \\
{\bf DDO71=KDG63}               &             & &             &              &              & &             &              &               &               \\
KDG63-3-1168 &20                & -4$\pm$30   &v&   6$\pm$2   & 0.4$\pm$0.3  &  -0.8$\pm$0.3& & 6.6$\pm$1.1 & 0.07$\pm$0.18&   6$\pm$0.7   & -1.2$\pm$0.1  \\
				&             & &             &              &              & &             &              &               &               \\
{\bf Holmberg~IX}               &             & &             &              &              & &             &              &               &               \\
HoIX-4-1038 & 24                & 40$\pm$10   & &5.7$\pm$1.9  & 0.2:         & -2.3$\pm$0.3 &v& 0.1$\pm$0.05& 0.05$\pm$0.03&  0.1$\pm$0.004& -0.4: \\
				&             & &             &              &              & &             &              &               &               \\
{\bf UGCA86}                    &             & &             &              &              & &             &              &               &               \\
13 & 23                         &  70$\pm$30  & &  3$\pm$3    & 0.0:         & -2.0$\pm$0.5 &v& 0.1$\pm$0.07& 0.17$\pm$0.08&   0.1$\pm$0.01& -0.64$\pm$0.11\\
20 & 24                         &  61$\pm$30  &v&0.5$\pm$0.2  & 0.0:         & -0.2$\pm$0.3 & & 0.8$\pm$0.4 & 0.66$\pm$0.3 &   0.5$\pm$0.02& -0.21$\pm$0.07\\
22 & 22                         &  38$\pm$30  &v& 11:         & 0.0:         & -1.2$\pm$0.2 & & 6.8$\pm$0.8 & 0.07$\pm$0.18&   13.5$\pm$1.2& -1.23$\pm$0.04\\
32 & 44                         &  60$\pm$30  &v& 10:         & 0.1:         & -1.7$\pm$0.2 & & 5.8$\pm$1.4 & 0.05$\pm$0.19&   10.8$\pm$0.9& -2.28$\pm$0.03\\
				&             & &             &              &              & &             &              &               &               \\
{\bf  M33}                      &             & &             &              &              & &             &              &               &               \\
{\bf 2},MKKSS27   &      96     & -168$\pm$18 &v& 2.5$\pm$1.6 & 0.0$\pm$0.2  &  0.1$\pm$0.3 & & ....        & ....         &   1.5$\pm$0.01& -0.10$\pm$0.01\\
{\bf 3},CBF99     &     36      & -240$\pm$14 &v& 0.7$\pm$0.2 & 0.3$\pm$0.2  & -0.1$\pm$0.2 & & 0.9$\pm$0.4 & 1.00:        &   0.7$\pm$0.02& -0.28$\pm$0.02\\
{\bf 4},KK6       &     136     & -269$\pm$4  &v&  10$\pm$2   & 0.2$\pm$0.2  & -1.0$\pm$0.2 & & 7.2$\pm$1.7 & 0.05$\pm$0.13&   6.7$\pm$0.05& -1.05$\pm$0.01\\
{\bf 5},CBF120    &     30      & -266$\pm$14 &v& 0.9$\pm$0.2 & 0.0$\pm$0.3  & -0.6$\pm$0.3 & & ....        & ....         &   1.3$\pm$0.02& -0.65$\pm$0.04\\
{\bf 6},CBF98     &     103     & -220$\pm$9  &v&  11$\pm$1   & 0.4$\pm$0.2  & -1.3$\pm$0.1 & & 6.5$\pm$0.6 & 0.05$\pm$0.13&   3.6$\pm$0.05& -0.87$\pm$0.01\\
{\bf 7},MKKSS33   &     32      & -193$\pm$16 &v& 1.2$\pm$0.7 & 0.3$\pm$0.3  &  0.2$\pm$0.2 & & ....        & ....         &   0.8$\pm$0.02& 0.16$\pm$0.02 \\
{\bf 8},CBF119    &     22      & -242$\pm$39 &v& 0.95$\pm$0.3& 0.2$\pm$0.3  & -0.5$\pm$0.4 & & 2.1$\pm$0.8 & 0.40$\pm$0.30&   1.0$\pm$0.05& -0.14$\pm$0.07\\
{\bf 9},CS R12,CBF116&     111  & -242$\pm$20 &v&  11$\pm$2   & 0.1$\pm$0.01 & -0.6$\pm$0.2 & & 9.6$\pm$0.5 & 0.05$\pm$0.10&   10.$\pm$0.30& -0.73$\pm$0.01\\
{\bf 10},CS U89, CBF56& 40      & -152$\pm$20 &v& 1.5$\pm$0.5 & 0.5:         & -1.3$\pm$0.2 & & 2.8$\pm$0.9 & 0.08$\pm$0.14&   0.6$\pm$0.03& -0.84$\pm$0.04\\
{\bf 11},CS U78, MKKSS42 & 38   & -201$\pm$14 &v& 0.3$\pm$0.15& 0.0:         & -1.7$\pm$0.4 & & 1.0:        & 0.05$\pm$0.13&   0.4$\pm$0.02& -0.04: \\
{\bf 12},CS U82, MKKSS44 & 44   & -163$\pm$15 &v& 0.3$\pm$0.07& 0.0$\pm$0.26 & -0.2$\pm$0.1 & & 0.2$\pm$0.08& 0.19$\pm$0.12&   0.2$\pm$0.01& 0.20$\pm$0.05 \\
{\bf 13},CS U83, MKKSS41 & 19   & -239$\pm$13 &v& 0.7$\pm$0.3 & 0.3:         & -0.3$\pm$0.3 & & 0.8$\pm$0.6 & 1.26:        &   0.5$\pm$0.03& -0.14$\pm$0.09\\
{\bf 1},CBF129    &     20      & -196$\pm$11 &v&   1:        &  ....        & -0.7:        & & ....        & ....         &   1.5$\pm$0.3 & -1.7:         \\
{\bf 14},CBF118   &     17      & -157$\pm$13 &v& 0.9:        & 0.5:         &  0.2:        & & 3.5: & 0.49$\pm$0.27 &  1.1$\pm$0.1 & -0.33$\pm$0.16\\
{\bf 15},CS U83, CBF152& 18     & -326$\pm$10 &v& 1.5:        & ....         & -1.0:        & & 0.8$\pm$0.5 & 0.20:        &   1.1$\pm$0.09& -1.7:         \\
				&             & &             &              &              & &             &              &               &               \\
{\bf M31}                       &             & &             &              &              & &             &              &               &               \\
MKKSS61 &18                     & -113$\pm$27 &v&   8$\pm$2   & 0.3$\pm$0.2  & -0.8$\pm$0.2 & & 3.5$\pm$0.9 & 0.49$\pm$0.27&  2.6$\pm$0.3  & -0.50$\pm$0.06\\ 
MKKSS58 &14                     & -232$\pm$17 &v&  10$\pm$1   & 0.3$\pm$0.1  & -0.6$\pm$0.2 & &  6: & 0.20:        &  10.7$\pm$1.7 & -0.20$\pm$0.06\\
MKKSS72 &43                     & -13$\pm$21  &v& 0.5$\pm$0.2 &  0.0:        & -0.8$\pm$0.4 & & 0.1$\pm$0.05& 0.14$\pm$0.06&  0.4$\pm$0.15 & -1.0:         \\
\hline
\end{tabular}
\end{table*}

\begin{figure*}
\includegraphics[width=1.0\textwidth]{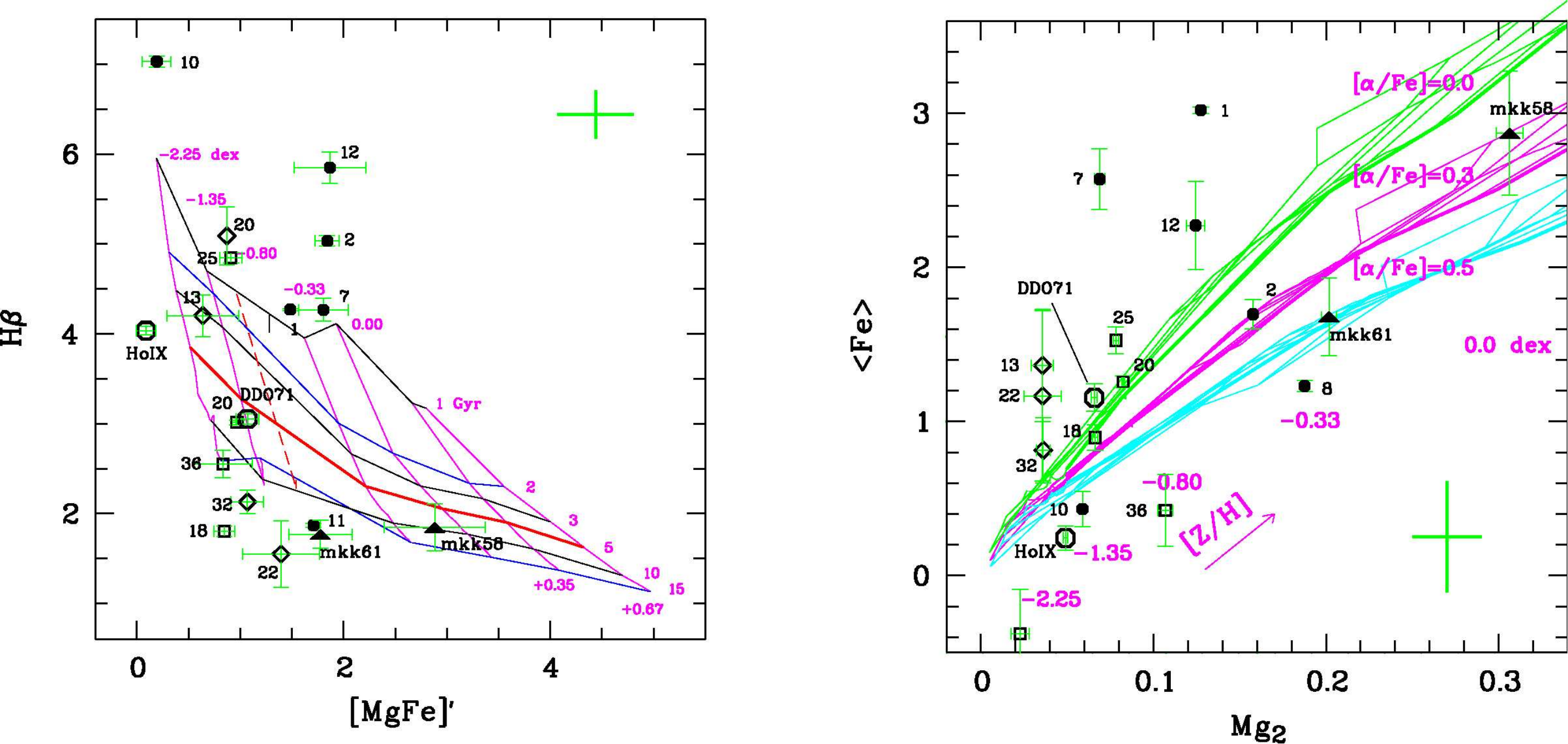}
\caption{Age -- metallicity (left panel), and metallicity -- \afe\ 
diagnostic plots (right panel). We use SSP model predictions of
Thomas et al. (2003, 2004). The cross in the corner of each panel
indicates the systematic calibration uncertainty to the Lick index
system. Symbols indicate GCs in different galaxies: M31 (triangles), M33
(dots), IC10 (open squares), UGCA86 (open lozenges), LSB dwarf galaxies
DDO71 and HoIX (large open circles).}
\label{DiagD}
\end{figure*}
\begin{figure*}
\includegraphics[width=0.5\textwidth, angle=-90]{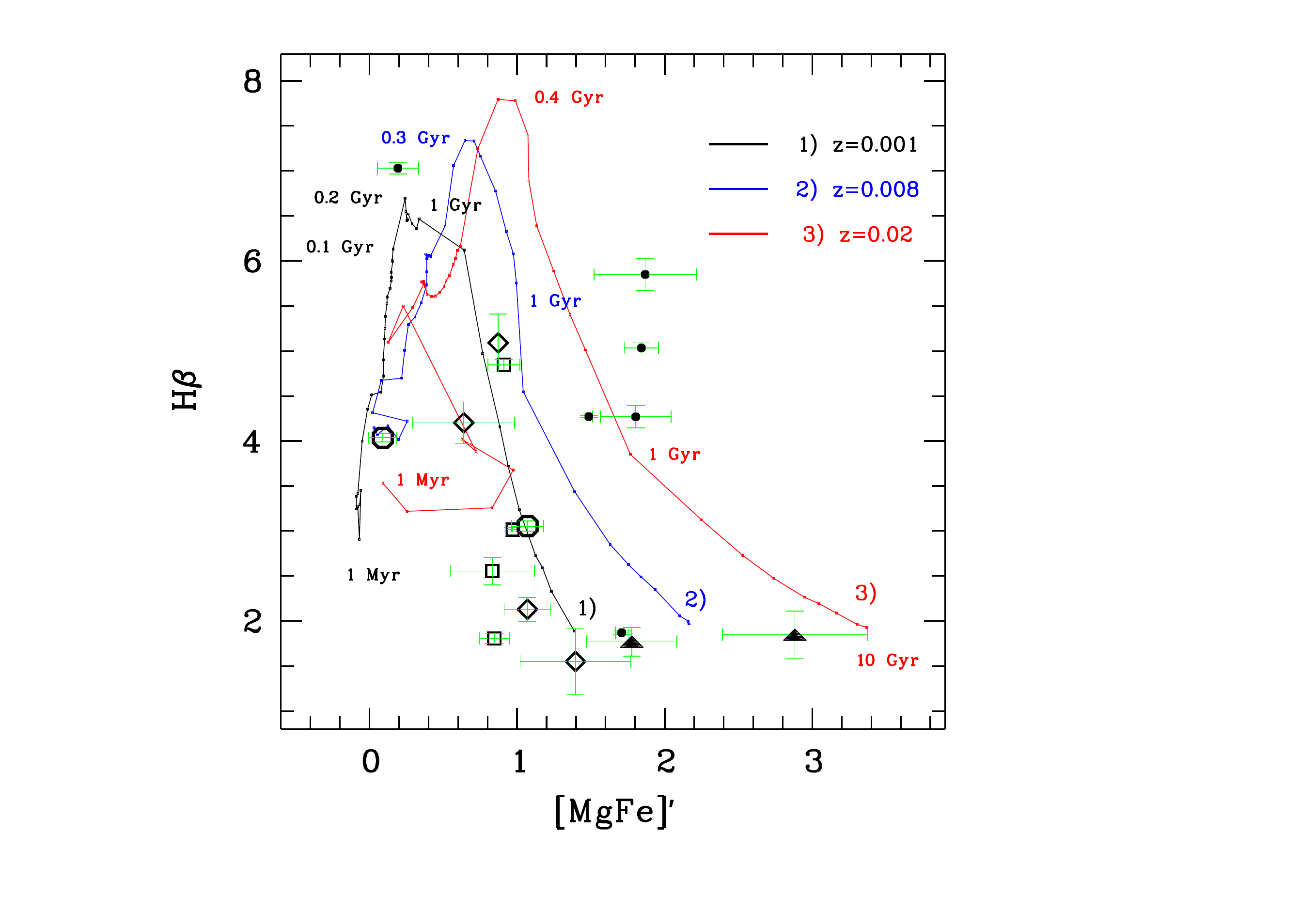}
\caption{H$\beta$ vs. \mgfe\ plot for the indices measured in the spectra 
of our program objects and in the model ones from Gonz\'{a}lez-Delgado
et al. (2005). The theoretical indices are shown for metallicities (Z)
1/20 solar, 2/5 solar, and solar. Points on the model sequences
correspond to the ages: 1 -- 9 Myr by steps of 1 Myr, 10 -- 95 by steps
of 5 Myr, 100 -- 900 by steps of 100 Myr , and 1 -- 10 Gyr by steps of 1
Gyr.}
\label{GD05}
\end{figure*}
\begin{figure*}
\includegraphics[width=1.0\textwidth]{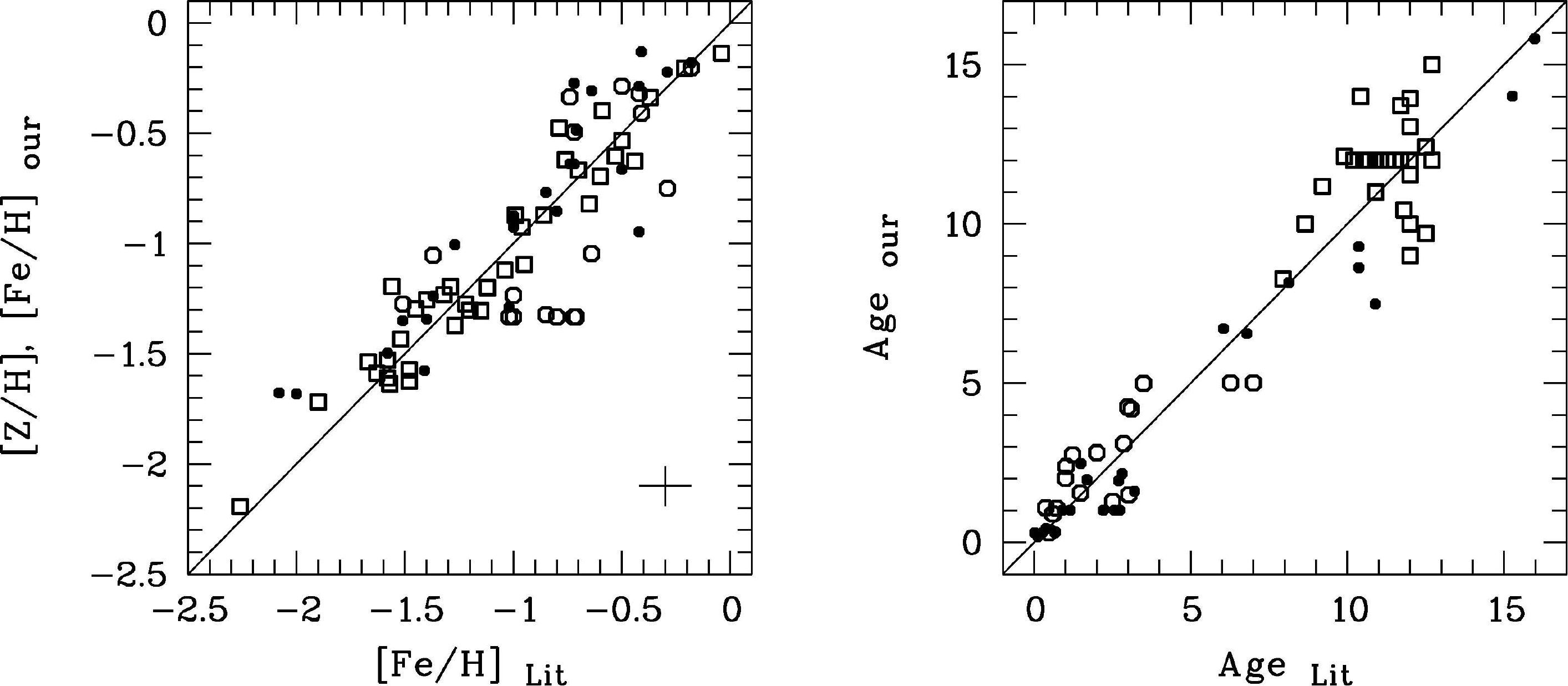}
\caption{Comparison between the literature ages and metallicities and 
those derived for Galactic GCs from the sample of Schiavon with the
models TMB03 (squares), and for LMC GCs with the models TMB03 (black
dots), and GD05 (open circles) (see text for details). The lines
indicate one-to-one relation. The cross in the corner of the left
panel indicates the average error of metallicity determinations. The
average accuracy of relative age determination is $\Delta t/t \approx 0.2$.}
\label{Compar}
\end{figure*}

\section{Determination of evolutionary parameters}
We obtained the evolutionary parameters of the GCs using three different
methods. The results are listed in Table~\ref{tab:evpar}.

The first approach (Sharina et al. 2006a, Sharina \& Davoust 2008)
allows one to derive age, [Z/H]\footnote{We use the standard definition,
$[X/Y]\!=\!\log(X/Y)\!-\!\log(X_{\sun}/Y_{\sun})$, where X and Y are
masses of specific elements. \zh\ is the overall metallicity. The
general relation between the metallicity Z and the iron content \feh\ is
$\log Z = \log (X/X_{\sun}) + \log Z_{\sun} + $[Fe/H], where X is the
hydrogen content, and all the symbols have their usual meaning (Bertelli
et al. 1994). The solar values are $X_{\sun}=0.70$ and $Z_{\sun}=0.02$.
The ratio $ X/X_{\sun}$ varies with the metallicity Z according to the
enrichment ratio $ \Delta Y/ \Delta Z$.} and $\alpha$-element abundance
ratio of GCs at once.
The $\chi^2$ minimization is carried out by comparing the
model Lick indices of Thomas et al., (2003, 2004; hereafter: TMB03), and
the measured ones weighted by their errors.
We interpolated linearly the model
indices within the three-dimensional space of model parameters:
age~$=0.5\!-\!15$ Gyr, $\zh=-2.25$ to $+0.35$~dex and
$\afe\!=\!0.0\!-\!0.5$ dex. The Age--\zh, \afe--\zh, and Age--\afe\
planes of the solution space for each GC with overplotted 67\%\, 95\%\,
and 99\%\ confidence contours are available on the SAO RAS
ftp-site\footnote{ftp://ftp.sao.ru/pub/sme/6m-spectr/ConfPlots/}. For
illustrative purposes we show our Lick index measurements on the
age--metallicity, and metallicity--\afe\ diagnostic diagrams
(Figure~\ref{DiagD}, and Figures~\ref{DiagD1} and \ref{DiagD2} in the
Appendix). Note that SSP models are only shown for solar
$\alpha$-element enhancements ($\afe = 0$~dex). Since, Balmer
line indices are primarily sensitive to age while \mgfe\ indices\footnote{
[MgFe]\arcmin~$= \left\{ \mbox{Mg}b \cdot (0.72 \cdot \mbox{Fe5270} +
0.28 \cdot \mbox{Fe5335})\right\}^{1/2}$} are sensitive to metallicity
and insensitive to \afe\ variations, a combination of diagnostic
diagrams (including Balmer, Mg, and Fe-line indices) are used for the
determination of evolutionary parameters (see Puzia et al. 2005a,b).
Unfortunately, not all GCs in our sample have measured \mgfe\ indices.
This is due to the shift of the sampled spectral range with the target 
position in the field of view of the SCORPIO spectrograph.

The second method is identical to the first one, except for the models
used. We measured the Lick indices in the model spectra of
Gonz\'{a}lez-Delgado et al. (2005) (hereafter: GD05), computed using
Geneva isochrones. The spectral library assumes solar abundance ratios
for all the elements, and covers the ages from 1 Myr to 10 Gyr, and
metallicities: 1/20 solar, 1/5 solar, 2/5 solar, solar, and twice solar.
For illustration, in Figure~\ref{GD05} we show a H$\beta$ -- \mgfe\ plot
with the indices measured in our spectra, and GD05 model ones. This
Figure shows that the model Balmer line sequences bend toward lower
index values for the SSPs younger than $\sim$300 Myr.
Because some of the indices become double-valued,
some indices may be interpreted as produced by younger clusters than if
one used the TMB03 models. There are two very young objects in our
sample. Indeed, the evolutionary parameters measured with the GD05
models for clusters HoIX-3-1038 and object \#13 in UGCA~86
(see Tab.~\ref{tab:evpar}) agree well with those obtained with the full
spectrum fitting method (see below), while in all other cases the
results obtained using the three methods are consistent with one
another.

To test the two aforementioned methods based on the analysis of Lick
indices we derive the evolutionary parameters for Galactic GCs using the
Lick indices measured in the spectra of Schiavon et al. (2005), and for
the Large Magellanic Clouds (LMC) GCs using the indices measured by
Beasley et al. (2002). The result of the comparison of ages and
metallicities obtained by us with the literature data is shown in
Figure~\ref{Compar}.~We transformed the metallicities, $Z$, obtained with
the GD05 models into \feh\ using the relation from Bertelli et al.
(1994): $\feh = 1.024\cdot\log Z+ 1.739$. The derived ages, metallicities,
and \afe\  are presented in Tables~\ref{lmc}, \ref{lmcGD05} and
\ref{Gal}. The literature metallicities were taken for Galactic
GCs from the catalog of Harris (1996), reference ages were taken from
Salaris \& Weiss (2002), and \afe\ from Pritzl et al. (2005) and Venn et
al. (2004). The evolutionary parameters for LMC GCs were summarised in
the paper of Beasley et al. (2002). The comparison shows that the
TMB03 models work very well within the age range $\sim$300 Myr -- 14
Gyr. It should be noted that the obtained mean $\alpha$-element ratio
for 41 Galactic GCs, $0.384\pm0.097$ dex, coincides with the
corresponding mean value known from high-resolution spectroscopic
studies for 13 GCs from this sample: $\afe = 0.366\pm 0.077$ dex 
(Venn et al. 2004, Pritzl et al. 2005). The GD05 models are
appropriate for analysing young GCs.

 Finally, we carry out an independent estimation of ages and
metallicities using the program ULySS (Koleva et al. 2009, and
references therein) and the Vazdekis SSP models (Vazdekis, 1999). To
provide full spectrum fitting with this method we analyse properly the
line-spread function (LSF) of the spectrograph, which is actually
different for each object in our case.

In general, the different methods give results that are in good
agreement with one another.
The few inconsistencies in the evolutionary parameters derived with the
three methods (see Tab.~\ref{tab:evpar} arise from cases of i) clusters
younger than $\sim$300 Myr; ii) non-solar \afe\, because the GD05 and 
Vazdekis models assume solar abundance
ratios for all the elements; iii) low S/N, where different SSP models
give different results. The presence of inhomogeneous foreground
emission, random and systematic errors may affect individual Lick index
measurements. The evolutionary parameters of low S/N objects are
uncertain and are marked by a colon in Table~\ref{tab:evpar}.
Additionally, the influence of hot horizontal-branch and blue straggler
stars on the integrated spectra may lead to artificially young ages for
old GCs.

The diagnostic plots illustrate general trends, in particular how
indices of different elements behave with respect to the SSP models, and
how the indices of GCs in different galaxies behave with respect to one
another.
The age-metallicity diagnostic diagrams (left panel in Fig.~\ref{DiagD}
and Fig.~\ref{DiagD1}) show that almost all GCs in our sample dwarf
galaxies are metal-poor ($\feh \le 0.8$ dex) and old (age $\ge 6$ Gyr).
GCs younger than $ \sim 1$ Gyr and of high metallicity ($\feh > 0.8$
dex) occur only in IC~10 and UGCA86, which are gas-rich dIrrs with clear
signatures of recent powerful starburst activity, and in the large
galaxies M33 and M31. Representatives of two populations of old GCs in
M31, metal-rich and metal-poor ones, are seen in the diagnostic plots.
Young clusters are generally more metal-rich.

Mg$_2$, which is sensitive to the $\alpha$-element ratio, is plotted
versus $\langle \mbox{Fe} \rangle $ index\footnote{ $\langle \mbox{Fe}
\rangle = (\mbox{Fe5270} + \mbox{Fe5335})/2$} in the right panel of
Figure~\ref{DiagD}, which shows that \afe\ is low for the GCs in our
sample dwarf galaxies. However, the difference between low and
high-\afe\ objects seems to be more marginal at low metallicities on
this diagram.

 The behaviour of the abundances of other elements with respect to
the SSP models is shown in Figure~\ref{DiagD2}. Since the CN$_2$ index
is sensitive to both N and C, and the over-abundance in Carbon (C) is
not evident from inspection of this diagram and diagnostic plots for
other indices sensitive to C, such as G4300 and C$_2$4668 (Fe4668), one
may conclude that there are signatures of over-abundance in Nitrogen (N)
of many GCs in IC10 and M33 (see CN$_2$ vs. \mgfe\ diagram). There are
no clear signatures of Calcium over-abundance for any galaxy in the
sample. High-resolution spectroscopy is needed to confirm these results.

\section{Results and discussion}
\subsection{Properties of the GC Systems}
\subsubsection{M33}

\begin{table*}
\scriptsize
\caption{Comparison of the measured radial velocities and evolutionary 
parameters for SCs in M33 (see Tables \ref{tab:evpar}) with literature data.
The superscript indices refer to the following data:
$^1$ Chistian \& Schommer (1983, 1988); $^2$ Chandar et al. (2002);
$^3$ Ma et al. (2001, 2002a,b, 2004a,b), $^4$ Sarajedini et al. (2007)}
\label{comparison}
\begin{tabular}{lcccrr} \\
\hline \hline
Cluster & V$_h^{lit}$  & $ log(age)^{our}$ & $\zh^{our}$ & $log(age)^{lit}$  & $\feh^{lit}$ \\
\hline
CBF 99  & -111$\pm21^2$& 9.0: & -0.7             & 9.1$^2$,9.11$^3$ & ....        \\
CBF 120 &  ....        & 9.3: & -0.8             & 6.6$^3$          & ....        \\
CS R14  & -224$\pm9^2$ & 10.0 &-1.2              & 9.6$^1$, 10.2$^2$, 9.11$^3$ & -1.5$^1$,-0.63$^3$, -1.0$^4$ \\
MKK 33  &  ....        & 9.08 &0.2               & 9.21$^3$         & -1.51$^3$     \\
CBF 119 &  ....        & 9.18 &-0.5              & 9.16$^3$         & -0.17$^3$     \\
CS U89  & -193$\pm16^2$& 9.18 &-1.5              & 9.25$^2$,8.01$^3$& -1.7$^3$ \\
CS U78  &  ....        & 8.9  & -1.5:            & 8.56$^3$         & ....         \\
CS U82  & -155$\pm27^2$& 8.84 &-1.2              & 8.96$^3$         & -1.7$^3$        \\
CS U83  &  ....        & 9.0  &-0.3              & 8.81$^3$         & ....         \\
CS U73  &  ....        & 9.18:&-1.0:             & 8.36$^3$         & ....         \\
CBF 118 & -154$\pm23^2$& 8.7: &-0.3              & 9.16$^3$         & 0.13$^3$       \\
\hline
\end{tabular}
\end{table*}

M33 is a relatively low-mass nucleated spiral ScII-III galaxy with disc
and halo components and without clear evidence of a bulge (van den Bergh
1999; Chandar et al. 2002; Brown 2009, and references therein). The
properties of SCs in M33 have been studied by many authors (see e.g.
Christian \& Schommer 1982, 1988; Brodie \& Huchra 1991; Sarajedini et
al. 1998, 2007; Chandar et al. 1999, 2001, 2002; Park \& Lee 2007). In
Table~\ref{comparison} we compare our determinations of cluster age and
metallicity with results from the literature. Our values agree with
previous estimates for the majority of objects, and match previous
spectroscopic results particularly well (e.g., results from Chandar et
al. 2002).

M33 is known to have formed clusters with a large range of ages and
metallicities. Sarajedini et al. (2000) suggested that the build-up of
the M33 halo extended over many Gyr, supported by the discovery in M33
of a genuine metal-poor intermediate-age GC  M33-C38, based on its
spectrum and color-magnitude diagram (Chandar et al. 2006). More
recently, Stonkut\.{e} et al. (2008) discovered an extended metal-poor
$\sim7$ Gyr old cluster M33-EC1.

Our sample clusters cover a wide range of ages and metallicities. The
mean value of \afe\ for 12 clusters with good spectra ($S/N \ge 20$) is
$\sim 0.2$~dex. There are five metal-rich SCs with ages younger than 1
Gyr, and a metal-rich one, MKKSS27 ({\bf \#2} in Tab.~\ref{tab:evpar}),
of intermediate age ($\sim$2.5 Gyr). One third of our sample appears to
be metal-poor ($\zh \le -0.8$ dex), with two young clusters, CS U89 and
CS U78 ({\bf \#10} and {\bf \#11} in Tab.~\ref{tab:evpar}).

\subsubsection{M31}

The GC system of our giant neighbour, M31, has been studied extensively
by many authors (see e.g the catalogue of Galleti et al. 2004, 2006 with
references and identifications of GCs found up to date). Huchra et al.
(1991) first obtained medium-resolution spectra of GCs in M31. In
distinction to the Milky Way, it has numerous intermediate-age and young
GCs. Morrison et al. 2004 found metal-poor thin-disc GCs, a strong
argument in favour of an undisturbed disc evolution, and of an
instantaneous inflow of weakly enriched gas.

We observed three GCs at the north-eastern end of the disc major axis,
nearer to the centre than the ''northern spur'' (Ferguson et al. 2002).
The galactocentric coordinates $X, Y$ (Huchra et al. 1991)\footnote{
$X\!=\!C_1\,\sin{(PA)}\,+\,C_2\,\cos{(PA)},$
$Y\!=\!-C_1\,\cos{(PA)}\,+\,C_2\,\sin{(PA)}$,
where $C_1\!=\![\sin(\alpha - \alpha_0) \cos{\delta}]$ and
$C_2\!=\![\sin \delta \cos \delta_0\!-\!\cos(\alpha \!-\! \alpha_0) \cos
\delta \sin \delta_0]$. $PA=37.7^\circ$ is the position angle for the
major axis of M31 (de Vaucouleurs 1958).} of the GCs MKKSS~61, 58 and 72
relative to the centre of M31 ($\alpha_0=00^h42^m44.3^s,
\delta_0=+41^\circ16\arcmin09\arcsec$, J2000.0, Crane et al. 1992) are
8.37, -1.44; 8.18, -1.37; and 8.9, -1.39 kpc, respectively. The radial
velocities of the two GCs MKKSS~61 and 72 agree with the mean thin disc
radial velocity of M31 in this region ($V_{hel} \sim -150$ km/s, see
Morrison et al. 2004, Chapman et al. 2006) with the rotation-corrected
velocity dispersion of the GC system of M31 at $|Y|= 1 - 3$ kpc, $\sigma
= 119$ km/s, estimated by Lee et al. (2008). The systemic radial
velocity of M31 is $V_{hel} = -300$ km/s (Mateo 1998). Additionally, the
velocity distribution of stars at the position of our sample clusters
(region D2 in Fig.~4, Chapman et al. 2006) indicates a clear disc-like
kinematics. The thick disc shows a slightly larger heliocentric velocity
in this field (see Fig.~2 in Chapman et al. 2006). Thus, our sample clusters
may be considered as belonging to the disc. The mean metallicity of the
three GCs is $\feh \sim -0.8$ dex. The mean age of the two old GCs is 9
Gyr. Our results agree well with those of Puzia et al. (2005a) for their
thin disc sub-sample. Comparison with the SF history known from stellar
photometry shows that these are two ordinary representatives of the old
stellar population of M31, because their ages and metallicities are
typical for many components of M31 (Brown, 2009): inner, outer, and
transition halo sub-components, and the stream. Our sample of GCs does
not show an overabundance in Calcium or Nitrogen, in distinction to the
majority of the M31 GCs studied by Puzia et al. (2005a).

\subsubsection{DDO71 = KDG63}

This dwarf spheroidal satellite of M81
contains only one relatively bright ($M_V = -7.2$)
GC near its optical centre (Karachentsev et al. 2000, KDG63-3-1168 in Sharina et al. 2005).
It has an intermediate age and a metallicity $\zh=-0.8 \pm0.3$,
but the S/N in the spectrum is quite low and the errors of
evolutionary parameters determination are large.
KDG63-3-1168 appear to be younger and
richer in metals compared to the many old
clusters found in M81 (Schroder et al., 2002; Ma et al. 2005, 2006,
2007; Georgiev et al. 1991b,c; Chandar et al. 2004).
Note that the mean metallicity of the GC is consistent with the mean
metallicity of red giant stars in DDO71 (\feh$=-1.17$) (Sharina et al., 2008).
The age of KDG63-3-1168 is similar to that of KK211-3-149,
the nucleus of a dSph galaxy in the Centaurus~A group (Puzia \& Sharina, 2008).

\subsubsection{Holmberg~IX}

There are several lines of evidence suggesting that HoIX is a tidal
dwarf galaxy, formed recently in tidal debris pulled out during the
latest major interaction between M81 and its satellites. These include
the large number of stars formed since the last major dynamical
interaction $\approx200$~Myr ago in the M81 system (Makarova et al.
2002; Sabbi et al. 2008; Williams et al. 2009), the relatively few RGB
stars and their unusual distribution suggesting that most of them belong
to M81 rather than HoIX (Sabbi et al. 2008). Fourteen star-cluster
candidates were found in HoIX based on high-resolution HST/WFPC2 images
(Sharina et al., 2005). Here, we obtained a spectrum of the brightest
blue cluster candidate from that list, which has $M_{V_0}=-9.05$ mag,
$(V-I)_0 =0.44$ mag, and an estimated mass of $\sim 10^6 M_{\sun}$.
Table~\ref{tab:evpar} shows that this cluster is slightly younger and
more metal-poor than the host galaxy itself.

\subsubsection{IC10}

IC10 is a dIrr satellite of M31, the nearest prototype starburst galaxy
(e.g. Thurow \& Wilcots, 2005), with a total baryonic mass ($M_{tot}$,
Tab.~\ref{tab:galaxies}) that is approximately an order of magnitude
lower than that of M33 (Corbelli 2003). IC10 is also known to have an
extended, old stellar halo (e.g. Tikhonov 1999, Demers et al. 2004), and
a giant HI-envelope extending far from the optical radius (Huchtmeier,
1979).

We found a number of SCs in IC10 (Tab.~\ref{ic10}) in addition to those
previously known (Hunter 2001).  Three GCs observed by us appear to be
metal-poor with ages in the range $5\!-\!12$ Gyr. We observed only one
young and metal-rich cluster ({\it 25}). It is similar to clusters {\it
CS U82} and {\it CS U83} in M33. The old and intermediate-age GCs appear
to be more metal-poor than the RGB stars in IC10, which have
$\overline{\feh} \sim -0.8$ dex (Dolphin et al. 2005, Battinelli et al.,
2007). Three of the sample clusters have near solar $\alpha$-element
ratios.

\subsubsection{UGCA~86}

This galaxy is actively forming stars, like IC10.
However this takes place only in two optically bright regions,
consistent with the maximum in HI distribution (Stil et al. 2005),
which extends far from the optical boundaries of the galaxy.
UGCA~86 covers a larger area than IC10 ($3.8\times2.7$ kpc$^2$ compared to
$1.1\times1.5$ kpc$^2$; see Karachentsev et al. 2004).
The centre of the dwarf galaxy is at a distance of 331~kpc\footnote{The
de-projected spatial separation between UA86 and IC342 (in Mpc) was calculated as
$R^2 = D_{UA86}^2\!+\!D^2_{I342}\!-\!2 D_{UA86}\!\cdot\!D^2_{I342}\!\cdot\!cos\Theta$,
where $\Theta$ is the angular separation in degrees, and
the distances to the galaxies are: $D_{UA86}=2.96$~Mpc, and
$D_{I342}=3.28$~Mpc (Karachentsev et al. 2006, 2004, 2002).} from the
centre of its spiral neighbor IC342.

There is a number of clusters and cluster candidates in UGCA86 (see
Tab.~\ref{ua86} and also Georgiev et al.~2009). Spectra for four of them
presented in this work suggest that two ancient clusters are metal-poor
like the old clusters observed in IC10. We also found one young
metal-rich cluster (\#20), and a young metal-poor one (\#13). The
objects do not have any enhancement of $\alpha$-elements relative to the
solar value.

\subsection{The Age--Metallicity Relation}

\begin{figure*}
\vspace{-0.7cm}
\begin{tabular}{p{0.5\textwidth}p{0.5\textwidth}}
\includegraphics[width=0.39\textwidth,angle=-90]{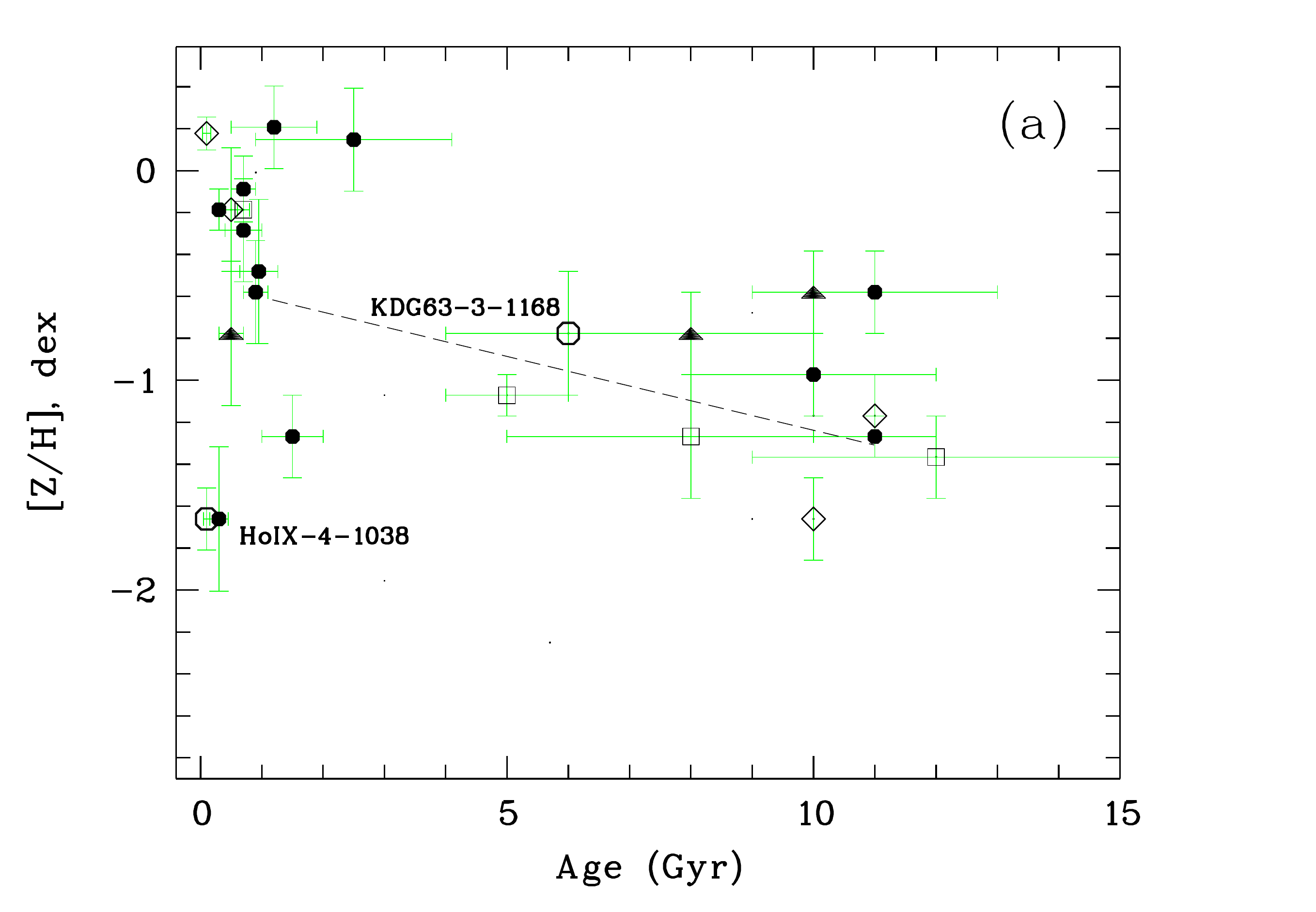}&
\includegraphics[width=0.39\textwidth,angle=-90]{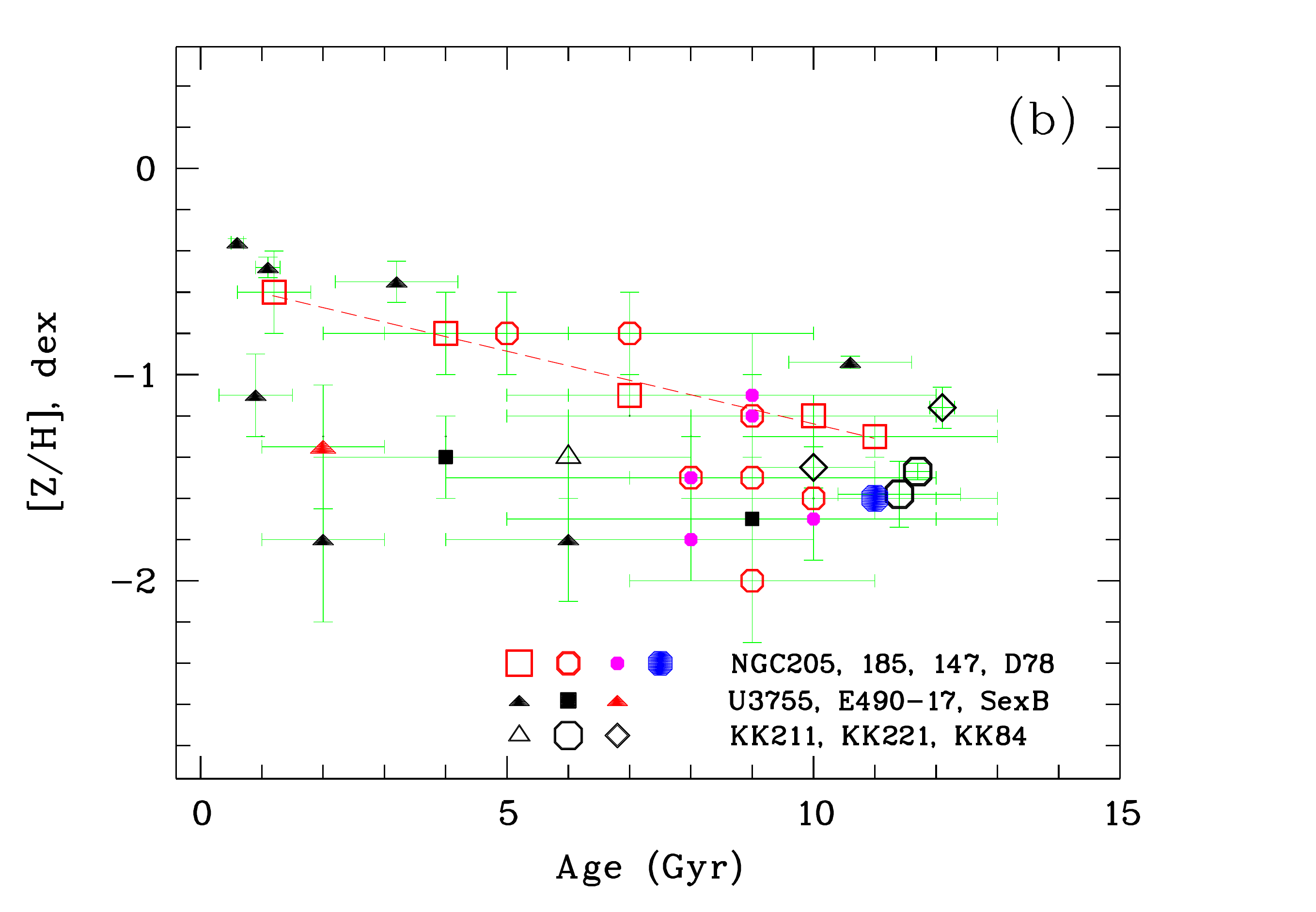}\\
\end{tabular}
\includegraphics[width=0.50\textwidth,angle=-90]{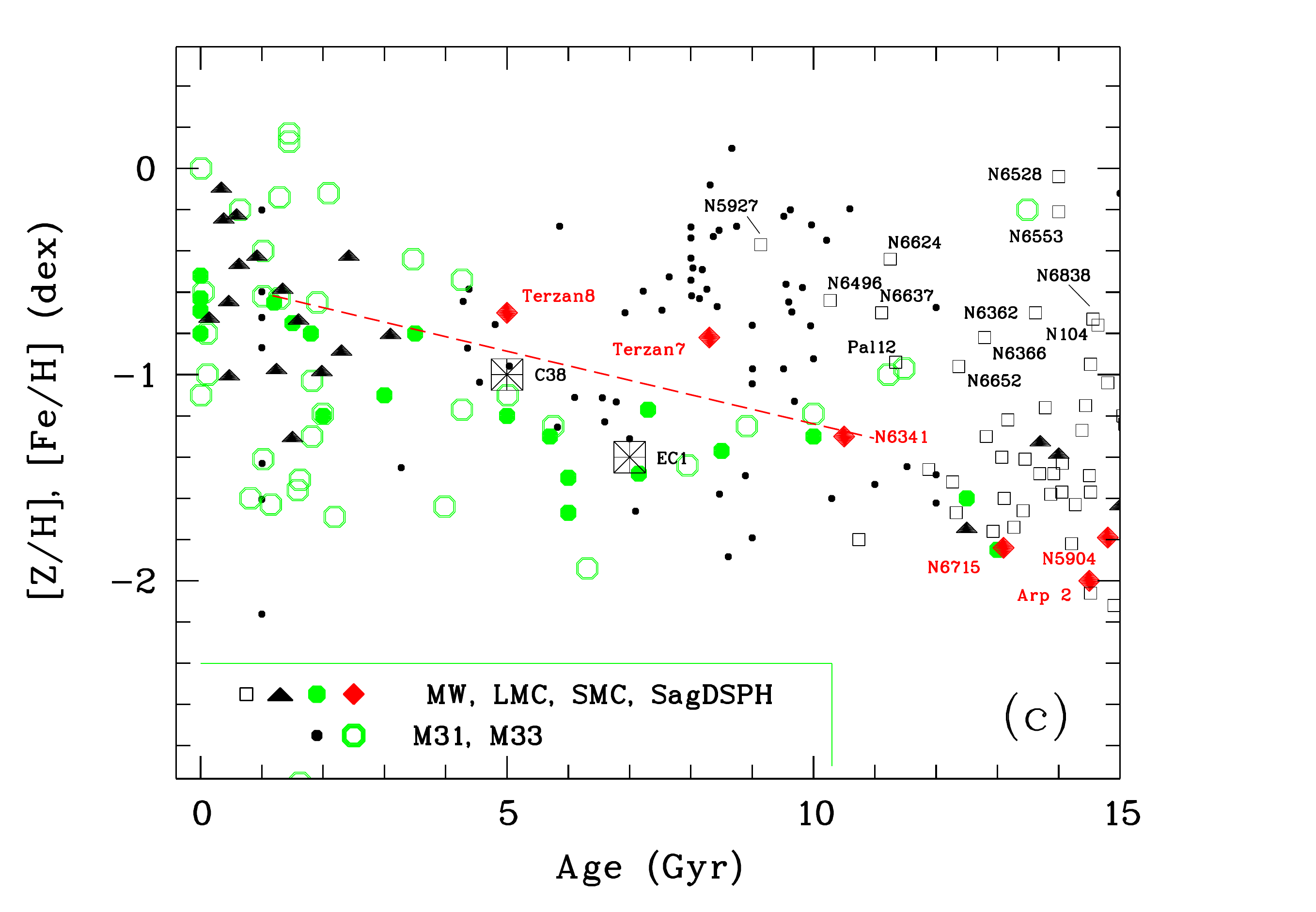}\\
\caption{Age-metallicity relation for (a) our sample of GCs 
(Tab.~\ref{tab:evpar}); (b) GCs in LSB dwarf galaxies and in the satellites of M31, NGC~147,
185, and 205; and (c) for LMC, SMC, Sagittarius dSph, and nearby spirals
M31 and M33 (see text for details). In the panel (a)
symbols are the same as in Figures~\ref{DiagD}, and \ref{DiagD2}.
Symbols designating GCs in different galaxies are marked in the 
plots (b) and (c). Intermediate-age and extended GCs in M33, M33-C38 and
M33-EC1, are indicated by crossed squares. The dashed line shows AMR for
GCs in NGC~205.}
\label{amr}
\end{figure*}

The age--metallicity relation (AMR) for our sample of GCs and for GCs in
other galaxies of different morphological types and masses in different
environments is shown in Figure~\ref{amr}. Since our observational
errors are large, the information retrievable from the plots is
approximate. However, it allows one to qualitatively compare the SF
histories of different galaxies. Since we did not target complete
samples of SCs for any galaxy, our data are supplemented by others from
the literature. The panel (a) represents our results from
Table~\ref{tab:evpar}. The panel (b) shows literature data for
LSB dwarf galaxies (Puzia \& Sharina 2008, Sharina et al. 2007, and
Sharina et al. 2003), and for satellites of M31, namely NGC~205, 185,
and 147 (Sharina et al. 2006a, Sharina \& Davoust 2008). The 
panel (c) shows the literature data for our Galaxy and M31 (Harris 1996, Puzia et
al. 2005a), LMC, SMC (Da Costa 1991, 2002, Da Costa \& Hatzidimitriou
1998, Beasley et al. 2002), Sagittarius dSph (Layden \& Sarajedini
2000, Forbes et al. 2004), M33 (Chistian \& Schommer 1988, Brodie \& Huchra 1991,
Sarajedini et al. 1998, 2007; Ma et al. 2001, 2002a,b,c; Chandar et al. 2002,
2006). The Sagittarius dSph GCs, the genuine intermediate-age
(Chandar et al. 2006) and extended (Stonkut\.{e} et al. 2008)
GCs in M33, and some Galactic GCs with $\feh\!>\!-1$~dex are labeled in
Figure~\ref{amr}. Note that all these Galactic GCs, except NGC~6362, have
very red horizontal-branch morphologies.

The AMR for GCs in NGC~205 is indicated by dashed lines in all the
panels. The SF histories of this galaxy and of the Sagittarius dSph
(Layden \& Sarajedini 2000) look similar. A simple closed-box model of
continuous SF fits well the data for NGC~205. The data for SMC follow
well the same model, but at a lower metallicity (Da Costa \&
Hatzidimitriou 1998). The enrichment histories of IC10 and M33 are
generally consistent with the model of continuous SF in the period
$\sim\!2.5\!-\!10$ Gyr with varying SF rate and different initial
metallicities.

While some galaxies or sub-systems form GCs continuously, others have a
single ancient powerful GC formation periods. These are our Galaxy and
dSphs (Fornax, NGC147, KK221, KK84, and DDO78). DDO71 and KK211
experienced powerful SF at intermediate ages. Interestingly, the GCs in
M31 roughly divide in two groups: old metal-rich objects, probably
belonging to the bulge of M31, and GCs of the disc. Surprising is the
presence of very metal-poor young and intermediate-age GCs in M31 and
M33 and their absence in some other relatively massive galaxies, for
example the Magellanic Clouds.

The most metal-poor {\it old} GCs are preferentially located in the 
halo of the MW, and in LSB dwarf galaxies: Sag dSph, UGCA86, 
HoIX, DDO78, KK221. The intermediate-age GCs in SagDSph, DDO71, 
KK221, and early-type dwarf satellites of M31, NGC205, and NGC185, are 
richer in metals than those in M33, SMC, IC10, and LSB dIrrs (UGC3755, 
ESO490-017, SexB, UGCA86, HoIX). The metallicity and age of 
KDG63-3-168 are comparable to the ones of GCs in the brighter early-type 
galaxies NGC205, and NGC185.

Both M31 and the MW formed GCs early in a wide range of metallicities.
However, our Galaxy started to form GCs at lower metallicity (and
possibly earlier) than M31. In distinction to the MW, there are many old
metal-rich SCs in M31. M33 shows a significant metallicity spread for
young SCs, similar to that of the more massive M31.

\subsection{Looking for a Link between Properties of the Galaxies and their GC Systems}

We now use our results for the clusters, in combination with information
compiled from the literature, to discuss the issue of the formation and
evolution of the clusters in each galaxy.

All dIrrs studied in our paper and M33 have sizable GC content and
enormous neutral hydrogen envelopes. The brightest SCs found by us 
have masses\footnote{The masses of GCs were estimated approximately
using the data on the visual V magnitudes from Tables~\ref{ic10} and
\ref{ua86}, distances and color excess values from
Table~\ref{tab:galaxies}, and typical mass-to-light ratios for GCs with
the ages 100 Myr -- 15 Gyr from Bruzual \& Charlot (2003).} $10^5-10^6
M_{\sun}$ in IC10, and $10^6-10^7 M_{\sun}$ in UGCA~86. If we only take
into account bright compact GCs, the number of GCs per unit galaxy mass
$M_G = 10^9 M_{\sun}$, $ T = \frac{N_{GC}(tot)}{M_G/10^9 M_{\sun}}$,
(Zepf \& Ashman, 1993) is $\sim$55 for IC10 and $\sim$40 for UGCA~86.
The unusual SC formation activity of both dwarf galaxies was probably
induced by the complex HI distribution in them with signatures of
interaction with the gas-rich galaxy M31 in the case of IC10 and of
infalling massive gaseous clouds in the case of UGCA~86 (Wilcots \& Miller 1998,
Stil et al. 2005).

The enormous gas contents of M31 and M33 and interaction between these
galaxies are likely reasons for active intermediate-age and young SC
formation. M33 has a huge warped HI disc, extending out to twice its
optical radius (Rogstad et al. 1976, Corbelli et al. 1989). Braun \&
Thilker (2004) discovered a faint HI bridge connecting M31 and M33 and
there is recent evidence for stellar tidal feature based on deep CFHT
photometry (McConnachie et al. 2009), which is evidence in favour of
mass transfer between the two galaxies.

Our sample dwarf galaxies are located approximately at the same distance
from their nearest massive neighbour (see Tab.~\ref{tab:galaxies})
$D_{MD}\sim 200-300$ kpc, except for the tidal dIrr HoIX ($D_{MD}
\sim70$ kpc).  The distribution of nearby dSphs ($D< 10$~Mpc) according
to their projected separation from the brightest galaxy in the group is
well fitted by an exponential with a scale length of $D_{MD}\sim200$ kpc
(Karachentsev et al., 2005). Thus, our sample galaxies lie approximately
near this borderline, within which the probability to find dSph is high.

dSphs are old and metal-poor stellar systems, as evidenced by their
color-magnitude diagrams (e.g. Grebel 1999). They have rather weak
metallicity gradients (Harbeck et al. 2001) in comparison to their
higher surface brightness counterparts. There are a few examples of GCs
located in the centres, or projected near the centres of dSphs with
metallicities similar to the mean metallicities of old stars in the host
galaxies: DDO78 in the M81 group (Sharina et al., 2003), KK211 and KK221 in the
Cen~A group (Puzia \& Sharina, 2008), and Hodge~I in NGC~147 (Sharina \&
Davoust, 2008). Interestingly, the nuclei of two other brighter
early-type satellites of M31, NGC~205 and 185, are much younger and more
metal-rich than the underlying stellar fields (e.g. Sharina et al. 2006a,
and references therein).

LSB early-type dwarf galaxies experience significant mass loss when
moving through the gaseous surroundings of nearby massive galaxies (e.g.
Gnedin, 2003). Observations of stars in the LG dSphs indicate that they
are dynamically "hot" systems in the sense that their internal stellar
velocity dispersion is larger than their mean velocity of rotation.
dSphs were more massive in the past, as evidenced from their SF
histories (Dolphin et al. 2005), the AMR (previous section), and the
high specific frequency of their GC systems (Miller \& Lots 2007).

\section{Conclusions}
We have obtained the evolutionary parameters of GCs in M31, M33, and,
for the first time, in the four low-mass galaxies IC10, UGCA86, DDO71,
and HoIX. Measurements of absorption-line indices in the well-known and
widely-used Lick system provide us with a suitable tool for studying the
ages and metallicities of these GCs.

In particular, we found a young and metal-poor massive GC in HoIX, the
tidal satellite of M81, the age of which is consistent with the age of
the host galaxy itself.

The central GC in DDO71, a dSph satellite of M81, appears to have an
intermediate age ($\sim 6$ Gyr) and metallicity ($\zh \sim -0.8$). It is
different from the many old clusters found in M81.

We observed SCs in a wide range of ages and metallicities in the two
actively star-forming dwarf irregular galaxies, IC10 and UGCA86. Their
GCs older than 1 Gyr are metal-poor, while their young clusters are
metal-rich. We found indications of continuous GC formation in IC10 and
an episodic one (old and young ($\la 1$~Gyr) periods) in UGCA~86.

The mean metallicity, age, and \afe\ obtained for the three GCs in the
disc of M31 ($\feh \sim -0.8$, age~$\sim 9$ Gyr, $\afe \sim 0.3$) agree
well with the values found for the thin disc GCs in M31 by Puzia et al.
(2005a). However our sample of GCs does not show overabundance in Calcium
and Nitrogen.

We obtained quite high S/N ratio spectra of twelve clusters in M33.
Among them are three old and metal-poor ($\zh \le -1.3$ dex), two young
and metal-poor, and seven young and metal-rich clusters. The presence of
metal-poor SCs in M33 in a wide range of ages may indicate an
instantaneous inflow of weakly-enriched gas into the galaxy, as was
found for M31 (Morrison et al. 2004).

The [$\alpha$/Fe] ratios of GCs were found to be higher on average in
giant than in dwarf galaxies.

We analysed the AMR for the globular clusters of our sample and others
from the literature. Although our knowledge about the evolutionary
parameters of the GCs is not complete, we may conclude that i) the
metallicity spread in GC systems is wider for larger galaxies; ii)
metal-rich clusters are young and preferentially found in galaxies more
massive than $\sim\!10^9 M_{\sun}$; iii)  intermediate-age globular
clusters in early-type dwarf galaxies are richer in metals than SCs
representing dynamically "cold" gas-rich environments in dIrrs; iv) the
AMR is special for each galaxy, and depends not only on its mass, but
also on some other factors, probably environmental conditions.

\vspace{0.2cm}
{\it Acknowledgements:}
The work was partly supported by grant RFBR~08-02-00627. We thank
Dr.~S.N.~Dodonov for supervising our observations. THP acknowledges
support through the Plaskett Fellowship at the Herzberg Institute of
Astrophysics of the National Research Council of Canada. Some of the
data presented in this paper were obtained from the Multimission Archive
at the Space Telescope Science Institute (MAST). STScI is operated by
the Association of Universities for Research in Astronomy, Inc., under
NASA contract NAS5-26555. Support for MAST for non-HST data is provided
by the NASA Office of Space Science via grant NAG5-7584 and by other
grants and contracts. Based on observations made with the NASA/ESA
Hubble Space Telescope, and obtained from the Hubble Legacy Archive,
which is a collaboration between the Space Telescope Science Institute
(STScI/NASA), the Space Telescope European Coordinating Facility
(ST-ECF/ESA) and the Canadian Astronomy Data Centre (CADC/NRC/CSA).

\appendix
\section{Photometry of clusters in IC10.}
\begin{table*}
\centering
\caption{Equatorial coordinates, magnitudes, colors, and central 
$V$ surface brightnesses of star clusters in IC10 found on the HST
images. Uncertain data due to inhomogeneous background in crowded
stellar fields are given without error bars and are marked by a colon.
In the first column we also indicate the objects from the paper of
Hunter (2001) (HX-X), and those which we consider to be common with the
list of Tikhonov \& Galazutdinova (2009) (TX).}
\scriptsize
\label{ic10}
\begin{tabular}{lclcccccr} \\ \hline
Sequence& RA(2000.0)DEC             &    $V$        &    $V-I$      & $ \mu_{V}$    &   $B-I$         \\
\hline
1=T2  & 00 19 57.84,+59 19 51.8  & 21.48$\pm$0.04&  0.84$\pm$0.07& 21.28$\pm$0.11&  1.86$\pm$0.07\\ 
2     & 00 19 58.08,+59 19 46.7  & 22.50$\pm$0.25&  1.75$\pm$0.25& 22.15$\pm$0.33&  3.20$\pm$0.30\\ 
d2    & 00 19 58.20,+59 19 40.3  & 22.63:        &  1.39$\pm$0.04& 21.75$\pm$0.16&  3.20$\pm$0.08\\ 
3     & 00 19 58.89,+59 19 59.8  & 22.45$\pm$0.30&  0.95$\pm$0.30& 21.60$\pm$0.14&  2.00$\pm$0.30\\ 
4=T4  & 00 20 00.06,+59 19 58.2  & 19.95$\pm$0.05&  1.44$\pm$0.05& 20.51$\pm$0.05&  2.90$\pm$0.04\\ 
d1    & 00 20 00.14,+59 19 44.1  & 21.86:        &  0.98$\pm$0.04& 20.65$\pm$0.23&  2.18$\pm$0.14\\ 
5     & 00 20 01.12,+59 19 30.1  & 22.80$\pm$0.07&  0.96$\pm$0.06& 20.92$\pm$0.13&  2.21$\pm$0.09\\ 
6=T5  & 00 20 01.94,+59 19 45.5  & 20.68$\pm$0.24&  1.28$\pm$0.09& 21.10$\pm$0.40&  2.64$\pm$0.15\\ 
7=T6  & 00 20 02.06,+59 20 04.6  & 21.35:        &  1.32$\pm$0.17& 20.6:         &  2.85$\pm$0.05\\ 
8=T7  & 00 20 02.34,+59 19 08.3  & 21.65$\pm$0.03&  1.82$\pm$0.06& 21.51$\pm$0.20&  3.80$\pm$0.16\\ 
d4    & 00 20 02.57,+59 19 31.1  & 22.96$\pm$0.07&  1.88$\pm$0.30& 21.57$\pm$0.22&  3.57$\pm$0.40\\ 
9     & 00 20 02.70,+59 20 08.3  & 22.55$\pm$0.07&  1.00$\pm$0.07& 21.84$\pm$0.13&  1.99$\pm$0.08\\ 
10=T8 & 00 20 03.25,+59 18 50.6  & 22.14:        &  1.99$\pm$0.08& 22.11$\pm$0.12&     ...       \\ 
11=T9 & 00 20 04.41,+59 18 35.0  & 21.45$\pm$0.20&  1.85$\pm$0.10& 22.25$\pm$0.20&  3.64$\pm$0.19\\ 
d5    & 00 20 04.54,+59 19 42.7  & 22.03:        &  0.93$\pm$0.07& 21.75$\pm$0.05&  2.52$\pm$0.08\\ 
12=T10& 00 20 05.76,+59 18 26.0  & 20.11$\pm$0.10&  1.88$\pm$0.05& 22.10$\pm$0.20&  3.83$\pm$0.05\\ 
d7     & 00 20 05.89,+59 19 04.7 & 23.12$\pm$0.10&  1.79$\pm$0.35& 22.08$\pm$0.74&  2.90$\pm$0.33\\ 
13=T11& 00 20 06.59,+59 19 22.7  & 20.87$\pm$0.11&  1.80$\pm$0.30& 22.18$\pm$0.08&  3.64$\pm$0.13\\ 
14=T12& 00 20 06.90,+59 19 05.8  & 21.2:         &  2.00$\pm$0.15& 22.20$\pm$0.16&  3.51$\pm$0.24\\ 
15=T13& 00 20 07.37,+59 19 16.2  & 21.18$\pm$0.30&  1.61$\pm$0.14& 21.47$\pm$0.13&  3.04$\pm$0.08\\ 
16=T14& 00 20 07.56,+59 19 27.0  & 20.17:        &  2.71:        & 19.40$\pm$0.47&  4.00$\pm$0.50\\ 
17    & 00 20 07.69,+59 19 02.8  & 21.76:        &  2.09$\pm$0.11& 22.71$\pm$0.12&   ...         \\ 
18=T15& 00 20 09.72 +59 17 19.3  & 18.48:        &  1.71$\pm$0.31& 16.25$\pm$0.16&   ...         \\ 
d11=T16& 00 20 10.55,+59 18 21.3 & 21.31$\pm$0.06&  1.58$\pm$0.11& 21.05$\pm$0.12&  3.17$\pm$0.16\\ 
d10   & 00 20 10.93,+59 18 25.4  & 20.57$\pm$0.2 &  1.46$\pm$0.08& 22.30$\pm$0.21&  2.90$\pm$0.17\\ 
d9=T18& 00 20 11.54,+59 18 50.5  & 19.16:        &  1.41$\pm$0.27& 20.84$\pm$0.25&  2.68$\pm$0.33\\ 
19=T19& 00 20 12.43,+59 19 16.5  & 19.03:        &  1.65$\pm$0.25& 18.97$\pm$0.16&  3.00$\pm$0.40\\ 
20=T20& 00 20 12.45,+59 17 28.0  & 17.70:        &  1.86$\pm$0.15& 16.56$\pm$0.15&  3.25$\pm$0.11\\ 
21    & 00 20 13.44,+59 20 16.1  & 21.33$\pm$0.05&  1.87$\pm$0.10& 22.12$\pm$0.13&  3.32$\pm$0.28\\ 
22=T21& 00 20 13.78,+59 21 14.8  & 21.98$\pm$0.20&  1.51$\pm$0.09& 21.55$\pm$0.15&  2.80$\pm$0.12\\ 
d27   & 00 20 14.30,+59 17 31.5  & 20.90:        &  2.30$\pm$0.37& 22.56$\pm$0.11&  4.19$\pm$0.55\\ 
d13   & 00 20 15.03,+59 19 06.2  & 22.11:        &  1.44$\pm$0.04& 20.94$\pm$0.11&  2.94$\pm$0.18\\ 
23=T22& 00 20 15.39,+59 19 50.9  & 22.22$\pm$0.15&  1.24$\pm$0.15& 21.44$\pm$0.22&  2.44$\pm$0.15\\ 
24=T23& 00 20 17.20,+59 17 01.1  & 19.52$\pm$0.14&  1.73$\pm$0.06& 20.06$\pm$0.18&  3.52$\pm$0.07\\ 
25=T24& 00 20 17.24,+59 17 45.3  & 17.68$\pm$0.02&  1.50$\pm$0.06& 15.76$\pm$0.08&  2.93$\pm$0.08\\ 
26=T25& 00 20 17.37,+59 16 56.1  & 20.81$\pm$0.08&  1.15$\pm$0.15& 20.48$\pm$0.09&  2.39$\pm$0.20\\ 
d12=T26& 00 20 17.71,+59 19 17.6  & 20.58$\pm$0.05&  1.13$\pm$0.11& 21.27$\pm$0.10&  2.47$\pm$0.22\\ 
d28=T27& 00 20 17.79,+59 17 46.1  & 18.80:        &  0.74$\pm$0.06& 15.94$\pm$0.26&  1.81$\pm$0.05\\ 
d21=T28& 00 20 17.90,+59 17 02.5  & 19.45$\pm$0.03&  1.50$\pm$0.14& 21.24$\pm$0.16&  2.79$\pm$0.12\\ 
27=T29& 00 20 17.91,+59 19 49.5  & 19.94$\pm$0.09&  1.51$\pm$0.13& 21.13$\pm$0.09&  2.77$\pm$0.16\\ 
d19=T30 & 00 20 18.32,+59 17 58.3& 19.74$\pm$0.2 &  1.53$\pm$0.20& 19.67$\pm$0.16&  2.50$\pm$0.29\\ 
28=T31& 00 20 18.44,+59 18 23.3  & 20.68$\pm$0.10&  1.78$\pm$0.11& 21.30$\pm$0.14&  3.32$\pm$0.15\\ 
d18=T33 & 00 20 18.93,+59 18 08.9& 20.08$\pm$0.11&  1.25$\pm$0.30& 20.69$\pm$0.10&  2.32$\pm$0.17\\ 
29=T34& 00 20 19.33,+59 17 30.6  & 19.85$\pm$0.3 &  1.64$\pm$0.10& 21.32$\pm$0.18&  3.22$\pm$0.1 \\ 
d14   & 00 20 19.82,+59 18 49.0  & 21.69$\pm$0.12&  1.43$\pm$0.30& 21.76$\pm$0.21&  2.78$\pm$0.30\\ 
d15=T35 & 00 20 20.07,+59 18 21.5& 20.51$\pm$0.11&  1.40$\pm$0.12& 20.70$\pm$0.14&  2.65$\pm$0.15\\ 
30=T36& 00 20 20.35,+59 18 37.4  & 18.31$\pm$0.2 &  1.48$\pm$0.14& 16.67$\pm$0.12&  2.53$\pm$0.20\\ 
31=T37& 00 20 20.93,+59 17 12.5  & 19.94$\pm$0.5 &  1.63$\pm$0.16& 20.57$\pm$0.12&  2.93$\pm$0.15\\ 
32=T38& 00 20 20.99,+59 18 58.9  & 21.35$\pm$0.15&  1.13$\pm$0.13& 21.06$\pm$0.12&  2.44$\pm$0.12\\ %
33(star?)& 00 20 21.53,+59 18 33.1  & 18.70$\pm$0.04&  1.27$\pm$0.09& 14.59$\pm$0.18&  2.75$\pm$0.07\\ 
d17   & 00 20 21.62,+59 18 25.3  & 20.23$\pm$0.20&  1.19$\pm$0.16& 20.53$\pm$0.13&  2.40$\pm$0.30\\ 
d16=T39 & 00 20 21.74,+59 18 22.4& 20.89$\pm$0.04&  1.28$\pm$0.07& 20.02$\pm$0.11&  2.73$\pm$0.24\\ 
d22=T42 & 00 20 23.10,+59 16 52.1  & 20.55$\pm$0.03&  1.43$\pm$0.09& 20.95$\pm$0.10&  2.70$\pm$0.30\\ 
H1-4  & 00 20 23.44,+59 17 51.2  & 18.61:        &  0.88$\pm$0.12& 21.18$\pm$0.19&  1.85$\pm$0.17\\ 
d24=T43& 00 20 23.91,+59 17 33.7  & 19.85:        &  1.31$\pm$0.14& 20.29$\pm$0.15&  2.48$\pm$0.20\\ 
H2-2  & 00 20 24.36,+59 19 10.1  & 21.10$\pm$0.2 &  0.95$\pm$0.11& 22.27$\pm$0.11&  2.20$\pm$0.16\\ 
H1-2  & 00 20 24.62,+59 18 11.9  & 18.62$\pm$0.07&  1.40$\pm$0.26& 17.12$\pm$0.14&  2.36 0.26    \\ 
H1-3  & 00 20 25.03,+59 17 38.9  & 18.82:        &  1.11$\pm$0.09& 20.52$\pm$0.08&  2.37$\pm$0.23\\ 
H1-1  & 00 20 25.17,+59 18 07.1  & 17.67:        &  1.31$\pm$0.31& 19.64$\pm$0.09&  2.34$\pm$0.33\\ 
H4-6 (star?)& 00 20 26.51,+59 16 36.3 & 18.93:   &  1.59$\pm$0.16& 14.57$\pm$0.15&  3.67$\pm$0.08\\ 
H4-3  & 00 20 26.70 +59 17 02.2  & 19.17:        &  0.89$\pm$0.12&               &  2.13$\pm$0.22\\ 
35=T51& 00 20 26.78,+59 19 46.9  & 22.30$\pm$0.25&  0.87$\pm$0.10& 22.52$\pm$0.06&  1.95 0.13    \\ 
H2-1  & 00 20 26.96,+59 18 16.9  & 20.29:        &  0.96$\pm$0.17& 21.81$\pm$0.16&  2.19$\pm$0.24\\ 
H4-7 (star?) & 00 20 27.58,+59 16 36.5& 18.86:   &  1.01$\pm$0.05& 14.07$\pm$0.17&  2.59$\pm$0.03\\ 
H4-4  & 00 20 27.60,+59 17 07.7  & 19.50:        &  0.97$\pm$0.04& 16.81$\pm$0.11&  2.20$\pm$0.04\\ 
36=T56& 00 20 29.56,+59 18 08.2  & 20.78(21.34)  &  1.57$\pm$0.05& 19.11$\pm$0.20&  3.38$\pm$0.04\\ 
d23=T57& 00 20 32.50,+59 17 12.8  & 20.76:        &  0.84$\pm$0.08& 21.59$\pm$0.18&  1.88$\pm$0.07\\ 
\hline
\end{tabular}
\end{table*}
\section{Photometry of star clusters in UGCA86.}

\begin{table*}
\centering
\caption{Equatorial coordinates, magnitudes, colors, and central $V$ surface
brightnesses of star clusters and candidates in UGCA86. In the first
column we indicate the objects common with the list of Georgiev et al.
(2009).}
\scriptsize
\label{ua86}
\begin{tabular}{lclccccr} \\ \hline
Sequence& RA(2000.0)DEC             &  $V$    &   $V-I$       & $ \mu_{V}$  \\ 
\hline
1    & 03 59 37.15,+67 07 00.0  & 23.14  & 1.41$\pm$0.04  & 21.39$\pm$0.05 \\ 
2    & 03 59 39.77,+67 06 49.0  & 22.59  & 1.33$\pm$0.03  & 21.09$\pm$0.04 \\ 
3    & 03 59 41.70,+67 08 10.5  & 22.00  & 0.21$\pm$0.02  & 19.72$\pm$0.03 \\ 
4=G20& 03 59 42.43,+67 08 53.1  & 22.74  & 1.51$\pm$0.06  & 21.43$\pm$0.03 \\ 
5    & 03 59 43.43,+67 06 01.3  & 22.42  & 1.58$\pm$0.01  & 21.23$\pm$0.04 \\ 
6    & 03 59 44.27,+67 07 14.6  & 22.04  & 1.21$\pm$0.02  & 19.77$\pm$0.05 \\ 
7    & 03 59 44.78,+67 08 33.8  & 22.30  & 1.28$\pm$0.04  & 20.97$\pm$0.05 \\ 
8    & 03 59 45.22,+67 06 57.5  & 22.82  & 1.33$\pm$0.05  & 21.33$\pm$0.06 \\ 
9    & 03 59 45.50,+67 09 13.7  & 23.43  & 1.28$\pm$0.03  & 22.04$\pm$0.06 \\ 
10   & 03 59 46.00,+67 07 39.3  & 22.71  & 1.44$\pm$0.13  & 21.36$\pm$0.04 \\ 
11   & 03 59 46.51,+67 07 26.2  & 22.80  & 1.26$\pm$0.06  & 21.24$\pm$0.05 \\ 
13   & 03 59 48.14,+67 08 19.6  & 19.60  & 1.28$\pm$0.01  & 18.21$\pm$0.02 \\ 
14=G17&03 59 48.79,+67 08 16.0  & 20.70  & 1.58$\pm$0.01  & 19.11$\pm$0.02 \\ 
15=G25&03 59 48.90,+67 08 30.0  & 22.02  & 1.21$\pm$0.06  & 19.87$\pm$0.09 \\ 
16   & 03 59 49.17,+67 08 21.8  & 22.20  & 1.03$\pm$0.15  & 20.80$\pm$0.06 \\ 
17=G28& 03 59 49.27,+67 08 40.0  & 22.64  & 1.62$\pm$0.06  & 20.17$\pm$0.05\\ 
18   & 03 59 49.28,+67 08 58.8  & 21.66  & 1.23$\pm$0.09  & 19.89$\pm$0.05 \\ 
19   & 03 59 49.34,+67 07 30.9  & 21.20  & 1.36$\pm$0.04  & 20.26$\pm$0.02 \\ 
20=G10&03 59 49.88,+67 06 49.1  & 19.31  & 1.61$\pm$0.01  & 16.58$\pm$0.01 \\ 
21   & 03 59 50.22,+67 08 14.6  & 22.50  & 1.37$\pm$0.04  & 20.51$\pm$0.04 \\ 
22=G29& 03 59 50.32,+67 08 37.4  & 19.28  & 1.76$\pm$0.01  & 18.49$\pm$0.01\\ 
23   & 03 59 50.93,+67 09 11.2  & 21.95  & 1.55$\pm$0.03  & 20.59$\pm$0.05 \\ 
24   & 03 59 50.97,+67 07 57.7  & 22.51  & 1.42$\pm$0.04  & 21.13$\pm$0.05 \\ 
25   & 03 59 51.06,+67 08 46.3  & 23.00  & 1.41$\pm$0.07  & 21.54$\pm$0.03 \\ 
26   & 03 59 51.32,+67 08 42.7  & 23.32  & 1.33$\pm$0.06  & 21.08$\pm$0.06 \\ 
27   & 03 59 52.45,+67 08 41.9  & 23.21  & 1.54$\pm$0.20  & 21.87$\pm$0.09 \\ 
28   & 03 59 52.63,+67 08 11.6  & 21.77  & 1.20$\pm$0.01  & 19.48$\pm$0.04 \\ 
29   & 03 59 53.75,+67 08 15.5  & 22.37  & 1.34$\pm$0.03  & 20.22$\pm$0.03 \\ 
30=G30?&03 59 53.91,+67 08 30.4  & 22.87  & 1.71$\pm$0.09  & 22.10$\pm$0.05\\ 
31   & 03 59 54.47,+67 07 51.8  & 21.39  & 1.29$\pm$0.10  & 19.50$\pm$0.03 \\ 
32   & 03 59 56.60,+67 06 11.6  & 18.70  & 1.48$\pm$0.02  & 16.63$\pm$0.03 \\ 
33   & 03 59 56.75,+67 07 38.4  & 21.03  & 1.36$\pm$0.03  & 18.42$\pm$0.02 \\ 
34   & 03 59 57.09,+67 07 37.0  & 22.88  & 1.12$\pm$0.05  & 20.53$\pm$0.04 \\ 
35   & 03 59 57.39,+67 06 12.4  & 20.49  & 1.04$\pm$0.17  & 17.82$\pm$0.13 \\ 
36   & 03 59 58.42,+67 06 10.4  & 21.37  & 1.63$\pm$0.07  & 19.10$\pm$0.04 \\ 
37=G27& 04 00 00.77,+67 07 36.5  & 22.07  & 1.65$\pm$0.05  & 20.94$\pm$0.04\\ 
\hline
\end{tabular}
\end{table*}
\section{Lick index measurements}
\begin{table*}
\tiny
\caption{Globular cluster indices ($\lambda \le 4531$\AA\ ) (first line)
corrected for zeropoints of transformation to the standard Lick system
and errors (second line indicated by the "$\pm$" sign) determined from
bootstrapping of the object spectrum.}
\label{lickind1}
\begin{tabular}{lrrrrrrrrrrrr} \\
\hline \hline
ID                 & H$\delta_{\rm A}$ & H$\gamma_{\rm A}$& H$\delta_{\rm F}$ & H$\gamma_{\rm F}$ & CN$_1$  & CN$_2$ & Ca4227 & G4300 & Fe4383 & Ca4455    \\
(S/N)              &(\AA)             & (\AA)          &  (\AA)              &    (\AA)             & (mag)    & (mag   & (\AA)  & (\AA) & (\AA)  & (\AA)     \\
 \hline
\noalign{\smallskip}
\noalign{\bf IC10}
\noalign{\smallskip}
18                     & 2.18             &  2.75          & 2.12                & 2.80                 & -0.025   & 0.015  & -0.22  & 0.03  & 2.14   & 0.66       \\
(35) \hskip 50pt $\pm$ & 0.06             &  0.07          & 0.07                & 0.07                 & 0.001    & 0.001  &  0.04  & 0.04  & 0.05   & 0.05       \\
20                     & ....             &  2.38          & 2.77                & 2.83                 & -0.034   & 0.011  &  0.15  & 0.93  & 1.24   & 0.91       \\
(42) \hskip 50pt $\pm$ & ....             &  0.03          & 0.03                & 0.03                 & 0.001    & 0.001  &  0.02  & 0.02  & 0.02   & 0.02       \\
25                     &     ....         &    6.06        &      ....           &          5.36        &  -0.077  &  -0.037& 0.43   &  1.41 &  1.24  &   0.92    \\
(75) \hskip 50pt $\pm$ &     ....         &     0.06       &      ....           &          0.06        &   0.001  &  0.001 & 0.04   &  0.04 &  0.05  &   0.05    \\
     36                & 3.08             &  1.71          & 3.85                & 3.50                 & 0.003    & 0.0002 &  1.37  & -0.23 & 1.43   & 1.46       \\
(20) \hskip 50pt $\pm$ & 0.18             &  0.18          & 0.19                & 0.19                 & 0.002    & 0.003  &  0.10  & 0.11  & 0.13   & 0.13       \\
\noalign{\smallskip}
\noalign{\smallskip}
DDO71-GC               &   2.40           &-0.47           & 2.08                &  1.94                & -0.068   & -0.006 & 0.35   & 3.15  & 1.71   & 2.18       \\
(20) \hskip 50pt $\pm$      &   0.07           & 0.07           & 0.07                &  0.07                &  0.001   & 0.001  & 0.03   & 0.04  & 0.04   & 0.05       \\
HoIX-1038              &   7.02           & 5.69           & 5.66                &  4.79                &  -0.069  & -0.106 & -0.10  & -0.86 & 1.06   & 0.60       \\
(24) \hskip 50pt $\pm$      &   0.06           & 0.06           & 0.06                &  0.06                &  0.001   & 0.001  & 0.03   & 0.03  & 0.04   & 0.04       \\
%
\noalign{\smallskip}
\noalign{\bf UGCA86}
\noalign{\smallskip}
13                     & 7.93             &  2.83          &  5.76               & 5.00                 & -0.139   & -0.057 & -0.36  & -1.64 & 0.66   & 0.32        \\
(23) \hskip 50pt $\pm$ & 0.27             &  0.28          &  0.29               & 0.29                 & 0.003    & 0.003  & 0.10   & 0.16  & 0.18   & 0.19        \\
20                     & 9.84             &  4.32          &  6.79               & 2.25                 & -0.156   & -0.049 &-0.44   & -0.64 & 0.83   & 0.41        \\
(24) \hskip 50pt $\pm$ & 0.30             &  0.30          &  0.30               & 0.30                 & 0.005    & 0.006  & 0.23   & 0.24  & 0.28   & 0.28        \\
22                     & -0.11            &  -4.01         &  1.62               & 0.37                 & -0.019   & -0.037 & 0.13   & 2.24  & 0.39   & 0.61        \\
(22) \hskip 50pt $\pm$ & 0.23             &  0.23          &  0.23               & 0.23                 & 0.005    & 0.007  & 0.15   & 0.16  & 0.26   & 0.26        \\
32                     & 3.64             &  -1.07         &  2.67               & 2.25                 & -0.087   & -0.049 & 0.29   & -0.64 & 0.83   & 0.41        \\
(44) \hskip 50pt $\pm$ & 0.11             &  0.11          &  0.11               & 0.11                 & 0.001    & 0.002  & 0.07   & 0.08  & 0.09   & 0.10        \\
\noalign{\smallskip}
\noalign{\bf M33}
\noalign{\smallskip}
1                      &  5.33           &   2.64        &       3.81          &       3.79            &  -0.101  &  -0.064 & -0.05  &-1.07  &  -0.41 &   0.42    \\
(20) \hskip 50pt $\pm$ &  0.06           &   0.06        &       0.06          &       0.06            &   0.0004 &   0.001 & 0.02   & 0.03  &  0.03  &   0.03    \\
2                      &  -1.63          &  -6.14        &     0.85            &     1.76              &   0.008  &   0.071 &   0.29 &  3.85 &  4.44  &   2.43      \\
(96) \hskip 47pt $\pm$&  0.02           &  0.02         &     0.02            &     0.02              &   0.0001 &   0.0002& 0.01   & 0.01  &  0.01  &   0.01      \\
3                      &  10.05          &   4.79        &     6.06            &     5.69              &  -0.180  &  -0.023 &   0.51 & -0.47 &  0.09  &   0.71      \\
(36) \hskip 50pt $\pm$ &  0.07           &  0.07         &     0.07            &     0.07              &   0.001  &   0.001 & 0.03   & 0.04  &  0.04  &   0.04      \\
4                      &          1.82   & -2.66         &     1.85            &     1.37              &   -0.046 &  -0.025 &   0.63 &   2.46&    0.99&   0.76  \\
(136) \hskip 47pt $\pm$&         0.02    &  0.02         &     0.02            &     0.02              &    0.0001&   0.0002&   0.01 &   0.01&   0.01 &   0.01  \\
5                      &          4.11   &  1.08         &     5.26            &     3.84              &   -0.152 &  -0.066 &  -1.17 &  -0.47&    0.38&   1.39  \\
(30) \hskip 50pt $\pm$ &         0.09    &  0.10         &     0.10            &     0.10              &    0.001 &   0.001 &   0.05 &   0.05&   0.07 &   0.07  \\
6                      &          3.15   & -2.21         &     3.05            &     1.76              &   -0.066 &  -0.020 &   0.42 &   1.89&    0.84&   1.11  \\
(103) \hskip 47pt $\pm$&         0.03    &  0.03         &     0.03            &     0.03              &    0.0002&   0.0004&   0.01 &   0.02&   0.02 &   0.02  \\
7                      &          3.60   &  2.52         &     5.08            &     5.19              &   -0.031 &   0.060 &   0.58 &   3.01&    4.78&   0.71  \\
(32) \hskip 50pt $\pm$ &         0.09    &  0.09         &     0.09            &     0.09              &    0.001 &   0.001 &   0.04 &   0.05&   0.05 &   0.06  \\
8                      &     ....        &  2.34         &     ....            &     3.56              &  ....    &  ....   & ....   &   2.93&    1.32&   0.58  \\
(22) \hskip 50pt $\pm$ &     ....        &  0.15         &     ....            &     0.15              &  ....    &  ....   & ....   &   0.04&    0.08&   0.09  \\
9                      &      -0.78      &  -5.87        &    -0.31            &    -0.72              &  -0.018  &  -0.010 &  0.54  &  3.24 &   2.15 &   0.78   \\
(111) \hskip 47pt $\pm$&      0.04       &  0.04         &     0.04            &     0.04              &   0.0003 &   0.0004&  0.01  &  0.02 &   0.02 &   0.02   \\
10                     &          6.85   &  4.43         &     4.67            &     5.30              &   -0.148 &  -0.092 &  -0.17 &  -0.32&   -1.99&   0.28  \\
(40) \hskip 50pt $\pm$ &         0.06    &  0.07         &     0.07            &     0.07              &    0.001 &   0.001 &   0.03 &   0.03&   0.04 &   0.04  \\
11                     &      11.89      &  8.70         &     7.43            &     7.61              &   -0.241 &  -0.168 &  0.11  &  -0.54&  -0.90 &   0.10  \\
(38) \hskip 50pt $\pm$ &      0.08       & 0.09          &     0.09            &     0.09              &    0.001 &   0.001 &  0.03  &   0.04&   0.05 &   0.05  \\
12                     &          9.96   &  6.11         &     6.27            &     5.99              &   -0.101 &  -0.113 &  -0.03 &  -2.99&   -0.91&   0.10  \\
(44) \hskip 50pt $\pm$ &         0.07    &  0.07         &     0.07            &     0.07              &    0.001 &   0.001 &   0.03 &   0.03&   0.04 &   0.04  \\
13                     &         8.83    &  6.56         &     6.47            &     5.78              &   -0.120 &   0.092 &   1.46 &   1.73&    1.54&   1.64  \\
(19) \hskip 50pt $\pm$ &        0.21     &  0.22         &     0.22            &     0.22              &    0.002 &   0.003 &  0.12  &   0.12&    0.14&   0.14  \\
14                     &         5.09    &  2.26         &     4.77            &     3.99              &   -0.024 &  -0.033 &   0.75 &   0.25&   -0.86&   0.42 \\
(17) \hskip 50pt $\pm$  &         1.04    &  1.06         &     1.07            &     1.08              &    0.010 &   0.014 &  0.51  &   0.56&    0.65&   0.68 \\
\noalign{\smallskip}
\noalign{\bf M31}
\noalign{\smallskip}
MKKSS61                &         1.19    & -1.89         &     2.59            &     1.82              &   -0.011 &  -0.006 &  0.99  &   2.50&   0.32 &  0.48 \\
(18) \hskip 50pt $\pm$ &         0.18    &  0.19         &     0.19            &     0.19              &    0.002 &   0.003 &  0.09  &   0.10&   0.12 &  0.12 \\
MKKSS58                &         1.36    & -5.76         &     0.77            &    -1.04              &    0.030 &  -0.069 &  0.93  &   1.82&   2.82 &  1.20 \\
(14) \hskip 50pt $\pm$ &         0.30    &  0.32         &     0.32            &     0.33              &    0.003 &   0.005 &  0.17  &   0.19&   0.21 &  0.21 \\
MKKSS72                &         8.38    &  5.62         &     5.93            &     5.91              &   -0.149 &  -0.117 &   0.13 &  -2.85&   1.93 &  0.35 \\
(43) \hskip 50pt $\pm$ &         0.05    &  0.05         &     0.05            &     0.05              &    0.0004&   0.001 &  0.02  &   0.03&   0.03 &  0.03 \\
\hline  \hline
\end{tabular}
\end{table*}

\begin{table*}
\tiny
\caption{Globular cluster indices ($\lambda \ge 4531$\AA\ ) (first line)
corrected for zeropoints of transformation to the standard Lick system
and errors (second line indicated by the "$\pm$" sign) determined from
bootstrapping of the object spectrum.}
\label{lickind2}
\begin{tabular}{llrrrrrrrrr} \\
\hline \hline
ID                 & Fe4531 & Fe4668 & H$\beta$ & Fe5015 & Mg$_1$   & Mg$_2$  & Mg$b$   & Fe5270 & Fe5335 & Fe5406 \\
(S/N)              & (\AA) & (\AA)  &  (\AA)   & (\AA)  & (mag)    & (mag)     & (\AA) & (\AA)  & (\AA)  & (\AA)  \\
 \hline
\noalign{\smallskip}
\noalign{\bf IC10}
\noalign{\smallskip}
18                     & 0.22   & -0.50 & 1.81     & 1.88    & 0.034    & 0.066  & 0.73   & 1.08    & 0.71   & 0.65       \\
(35) \hskip 50pt $\pm$ & 0.05   & 0.06  & 0.06     & 0.06    & 0.002    & 0.002  & 0.06   & 0.06    & 0.06   & 0.06       \\
20                     & 0.69   & 0.73  & 3.01     & 3.07    & 0.043    & 0.082  & 0.72   & 1.36    & 1.15   & 0.62       \\
(42) \hskip 50pt $\pm$ & 0.02   & 0.02  & 0.02     & 0.03    & 0.001    & 0.001  & 0.03   & 0.03    & 0.03   & 0.03       \\
25                     &  1.09  & 0.56  & 4.85     & 2.50    &  0.038   & 0.078  & 0.54   &  1.53   &   1.53 &     0.78 \\
(75) \hskip 50pt $\pm$ &  0.05  & 0.06  & 0.06     & 0.06    &  0.002   & 0.002  & 0.06   &  0.06   &   0.06 &     0.06 \\
     36                & -0.25  & 0.12  & 2.55     & 1.08    & 0.054    & 0.107  & 1.09   & 0.90    & -0.06  & 1.94       \\
(20) \hskip 50pt $\pm$ & 0.13   & 0.15  & 0.15     & 0.16    & 0.004    & 0.004  & 0.16   & 0.16    & 0.17   & 0.17       \\
\noalign{\smallskip}
\noalign{\smallskip}
DDO71-GC               & 0.54   & -0.12 & 3.05     & 3.15    & 0.044    & 0.066  & 1.03   & 1.07    & 1.24   & 0.74       \\
(20) \hskip 50pt $\pm$      & 0.05   &  0.06 & 0.06     & 0.06    & 0.002    & 0.002  & 0.06   & 0.06    & 0.06   & 0.06       \\
HoIX-1038              & 1.38   &  0.99 &  4.04    & 0.19    & -0.008   & 0.049  & 0.04   & 0.15    & 0.33   & 0.70       \\
(24) \hskip 50pt $\pm$      & 0.04   &  0.04 &  0.05    & 0.05    &  0.001   & 0.001  & 0.05   & 0.05    & 0.05   & 0.05       \\
%
\noalign{\smallskip}
\noalign{\bf UGCA86}
\noalign{\smallskip}
13                     & 0.85   & -1.59 &  4.20    &  -1.20  & 0.008    & 0.036  & 0.22   & 2.52    & 0.21   & ....        \\
(23) \hskip 50pt $\pm$ & 0.19   & 0.20  &  0.21    &  0.21   & 0.006    & 0.006  & 0.21   & 0.22    & 0.22   & ....        \\
20                     & 0.96   & 2.35  &  5.09    &  4.26   & ....     & ....   & 0.77   & 1.43    & -0.13  & ....        \\
(24) \hskip 50pt $\pm$ & 0.29   & 0.31  &  0.32    &  0.32   & ....     & ....   & 0.33   & 0.34    & 0.34   & ....        \\
22                     & 3.46   & -1.96 & 1.55     & 1.72    & 0.008    & 0.036  & 1.29   & 1.96    & 0.37   &  ....       \\
(22) \hskip 50pt $\pm$ & 0.24   & 0.25  & 0.26     & 0.26    & 0.010    & 0.010  & 0.26   & 0.27    & 0.27   & ....        \\
 32                    & -1.01  & -1.11 & 2.13     & 3.47    & -0.001   & 0.036  & 0.97   & 1.66    & -0.03  &  ....       \\
(44) \hskip 50pt $\pm$ & 0.11   &  0.13 & 0.13     & 0.14    & 0.004    & 0.004  & 0.15   & 0.15    & 0.15   &  ....       \\
\noalign{\smallskip}
\noalign{\bf M33}
\noalign{\smallskip}
1                      & -0.58  &    2.01 &   1.13  & 0.77  &   0.024   & 0.067  & 0.35   & -0.03   &   2.08 & .... \\
(20) \hskip 50pt $\pm$ & 0.04  &   0.04  &   0.04  & 0.05  &    0.001  & 0.001  &  0.05  &   0.05  &    0.06&  ....  \\
2                      &  4.63  &   2.88  &    4.27 &  8.95 &   0.034   & 0.127  &    0.73&    3.05 &    3.00& ....          \\
(96) \hskip 47pt $\pm$& 0.01  &    0.01 &    0.01 &  0.01 &    0.0004 & 0.0004 &  0.01  &   0.01  &   0.02 &  ....         \\
3                      &  1.73  &   0.88  &    5.03 &  3.76 &   0.044   & 0.158  &    2.61&    0.80 &    2.59&  ....         \\
(36) \hskip 50pt $\pm$ &  0.05  &    0.06 &    0.06 &  0.06 &    0.002  & 0.002  &  0.06  &   0.07  &   0.07 &  ....         \\
4                      &    1.71&     1.20&     1.88&  2.90 &   ....    & ....   & 1.59   &  ....   & ....   &  ....      \\
(136) \hskip 47pt $\pm$&   0.01&     0.01&     0.01&  0.02 &    ....   &  ....  & 0.02   &   ....   & ....   &  ....      \\
5                      &    0.95&     0.58&     3.59&  6.95 &   ....    & ....   & 2.68   & 3.11    & ....   &  ....      \\
(30) \hskip 50pt $\pm$ &    0.07&     0.08&     0.08&  0.09 &   ....    & ....   & 0.09   & 0.09    & ....   &  ....      \\
6                      &    1.55&    -0.12&     2.04&  2.73 &   ....    &  ....  & 1.48   & ....    & ....   &  ....      \\
(103) \hskip 47pt $\pm$&   0.02&     0.02&     0.02&  0.03 &    ....   &   .... & 0.03   &  ....    & ....   &  ....      \\
7                      &    3.97&     2.30&     4.65&  4.88 &   ....    &  ....  & 1.71   & 0.39    & ....   &  ....      \\
(32) \hskip 50pt $\pm$ &    0.06&     0.07&     0.07&  0.08 &   ....    &  ....  & 0.08   & 0.08    & ....   &  ....      \\
8                      &    1.57&    -1.46&     4.27&   4.09&    0.058  & 0.069  &   1.24 &    2.67 &    2.47&      0.59   \\
(22) \hskip 50pt $\pm$ &    0.10&     0.12&     0.12&   0.13&    0.003  & 0.003  &   0.13 &    0.14 &    0.14&      0.14   \\
9                      &   2.44 &    0.81 &    1.87 &  3.70 &   0.036   & 0.188  &   2.03 &   1.71  &   0.75 &     0.40     \\
(111) \hskip 47pt $\pm$&  0.02 &    0.02 &    0.02 &  0.02 &    0.001  & 0.001  &   0.02 &   0.03  &   0.03 &     0.04     \\
10                     &    0.89&     0.10&     5.84&  2.24 &   ....    &  ....  & ....   & ....    & ....   &  ....      \\
(40) \hskip 50pt $\pm$ &    0.04&     0.05&     0.05&  0.06 &   ....    &  ....  & ....   & ....    & ....   &  ....      \\
11                     &    1.08&    -0.08&    7.03 &  1.76 &  -0.044   & 0.059  &   0.08 &    0.46 &    0.41&  ....       \\
(38) \hskip 50pt $\pm$ &    0.05&     0.06&    0.06 &  0.07 &   0.002   & 0.002  &   0.08 &    0.08 &    0.08&  ....       \\
12                     &    2.11&     0.77&     6.14&  3.04 &   ....    &  ....  & ....   & ....    & ....   &  ....      \\
(44) \hskip 50pt $\pm$ &    0.04&     0.05&     0.06&  0.06 &   ....    &  ....  & ....   & ....    & ....   &  ....      \\
13                     &    3.16&     1.06&     5.85&   3.80&    0.069  & 0.124  &   1.65 &    1.90 &    2.64&      1.48   \\
(19) \hskip 50pt $\pm$ &    0.15&     0.17&     0.17&   0.19&    0.005  & 0.005  &   0.20 &    0.20 &    0.20&      0.20   \\
14                     &    2.90&     1.99&     3.17&  3.43 &   0.0207  & 0.1770 &    1.45&    1.76 &    0.69&   ....        \\
(17) \hskip 50pt $\pm$  &    0.70&     0.82&     0.83&  0.89 &   0.0232  & 0.0237 &    0.95&    0.97 &    0.98&   ....        \\
\noalign{\smallskip}
\noalign{\bf M31}
\noalign{\smallskip}
MKKSS61                &    2.83&     1.78&     1.77&  5.04 &    0.038  &  0.202 &   1.88 &   1.68  &   1.68 &  ....     \\
(18) \hskip 50pt $\pm$ &    0.13&     0.15&     0.16&  0.16 &    0.004  &  0.004 &   0.17 &   0.18  &   0.18 &  ....     \\
MKKSS58                &   -2.60&     1.97&     1.85&  6.23 &    0.063  &  0.306 &   2.80 &   3.08  &   2.66 &    0.81   \\
(14) \hskip 50pt $\pm$ &    0.23&     0.26&     0.26&  0.27 &    0.008  &  0.008 &   0.28 &   0.28  &   0.28 &    0.28   \\
MKKSS72                &   -0.15&    -0.34&     5.26&  2.99 &   ....    &  ....  &   1.36 & ....    & ....   &  ....         \\
(43) \hskip 50pt $\pm$ &    0.03&     0.04&     0.04&  0.05 &   ....    &  ....  &   0.05 &  ....   &  ....  &  ....         \\
\hline  \hline
\end{tabular}
\end{table*}

\begin{figure*}
\includegraphics[width=1.0\textwidth]{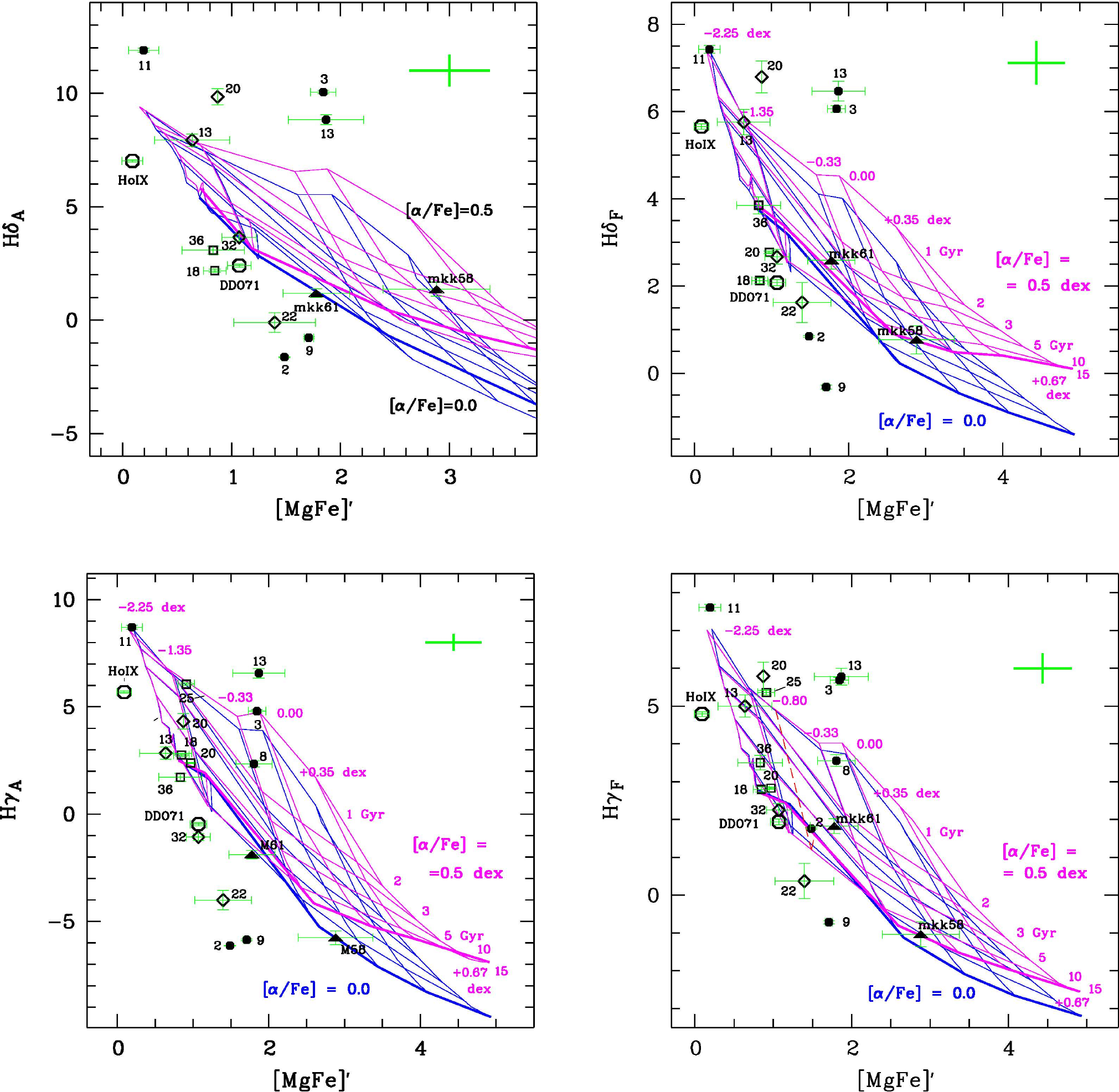}
\caption{Age -- metallicity diagnostic plots with high-order Balmer-line indices.
 We use SSP model predictions of
Thomas et al. (2003, 2004). The cross in the corner of each panel
indicates the systematic calibration uncertainty to the Lick index system.
In all panels, we plot model grids for $\afe=0.0$ and 0.5~dex. All ages, metallicities,
 and symbols are the same as in Figure~\ref{DiagD}.}
\label{DiagD1}
\end{figure*}
\begin{figure*}
\includegraphics[width=1.0\textwidth]{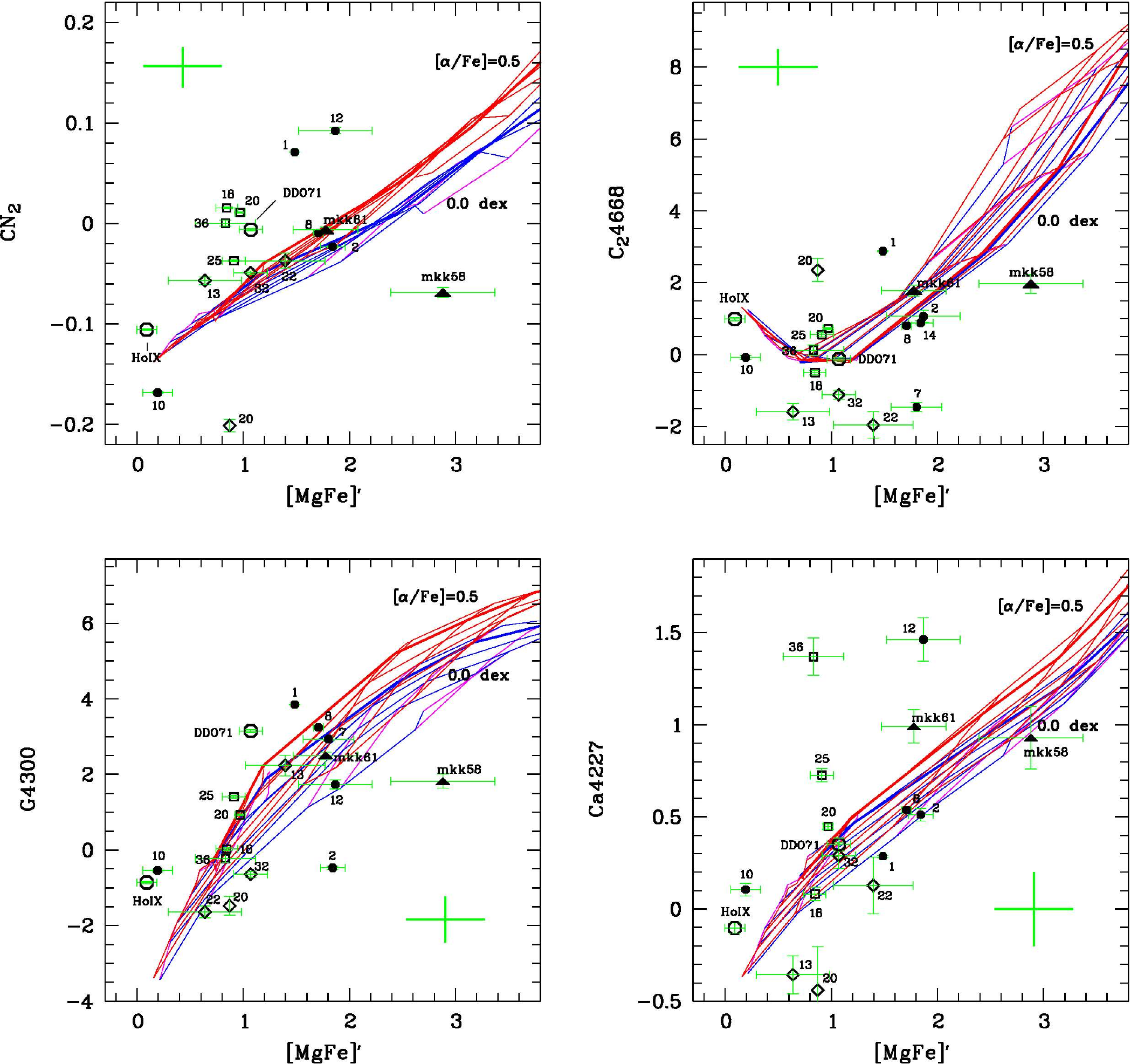}
\caption{Diagnostics plots of \mgfe\ versus indices sensitive to C, N, and Ca abundances.
Two model grids for $\afe=0.0$ and 0.5~dex are shown. All ages, metallicities, and 
symbols are as in Figure~\ref{DiagD}.}
\label{DiagD2}
\end{figure*}

\section{Estimated evolutionary parameters of globular clusters in our Galaxy and
in the Large Magellanic Cloud.}
\begin{table*}
\scriptsize
\caption{Evolutionary parameters of Galactic GCs from the sample of Schiavon
estimated by us using TMB03 models. In the last three columns the
corresponding literature data are shown. Literature \afe\ values were
taken from Pritzl et al. 2005 and Venn et al. 2004, the metallicities
were extracted from the catalog of Harris (1996), and the reference ages
were taken from Salaris \& Weiss (2002).
A colon is used when $\Delta\feh > 0.4$ dex, $\Delta$(age) $> 4$ Gyr, 
and $\Delta\afe > 0.35$ dex.}
\label{Gal}
\begin{tabular}{lcccccr} \\
\hline \hline
Cluster &  age$^{our}$ & $\afe^{our}$  & $\zh^{our}$ & $age^{lit}$  & $\afe^{lit}$  & $\feh^{lit}$  \\
	&   (Gyr)      &   (dex)       &    (dex)    &     (Gyr)    &   (dex)       &     (dex)     \\
\hline
NGC 104 & 12:            & 0.40$\pm$0.11  &  -0.62$\pm$0.18   &   10.7   &  0.29 &  -0.76 \\
NGC 1851& 11$\pm$3       & 0.31$\pm$0.13  &  -1.27$\pm$0.06   &    9.2   &       &  -1.22 \\
NGC 1904&  8$\pm$1       & 0.40$\pm$0.29  &  -1.64$\pm$0.22   &    7.9   &  0.35 &  -1.57 \\
NGC 2298& 10:            & 0.5:           &  -1.7:            &          &  0.34 &  -1.90 \\
NGC 2808& 12$\pm$2       & 0.30$\pm$0.24  &  -1.31$\pm$0.10   &   10.2   &       &  -1.15 \\
NGC 3201& 12$\pm$4       & 0.33$\pm$0.34  &  -1.53$\pm$0.21   &   11.3   &  0.22 &  -1.58 \\
NGC 5286& 10:            & 0.47$\pm$0.25  &  -1.54$\pm$0.12   &   12.0   &       &  -1.67 \\
NGC 5904& 11$\pm$4       & 0.39$\pm$0.16  &  -1.37$\pm$0.12   &   10.9   &       &  -1.27 \\
NGC 5927& 15:            & 0.39$\pm$0.09  &  -0.34$\pm$0.10   &          &       &  -0.37 \\
NGC 6121& 14:            & 0.5:           &  -1.3:            &   11.7   &       &  -1.20 \\
NGC 6171& 12$\pm$3       & 0.44$\pm$0.24  &  -1.12$\pm$0.10   &   11.7   &       &  -1.04 \\
NGC 6218& 12:            & 0.5:           &  -1.6:            &   12.5   &  0.35 &  -1.48 \\
NGC 6235& 14$\pm$1       & 0.26$\pm$0.29  &  -1.26$\pm$0.11   &   12.0   &       &  -1.40 \\
NGC 6254& 10:            & 0.46$\pm$0.24  &  -1.43$\pm$0.16   &   11.8   &       &  -1.52 \\
NGC 6266& 12$\pm$2       & 0.37$\pm$0.15  &  -1.20$\pm$0.07   &   12.0   &       &  -1.29 \\
NGC 6284& 12$\pm$3       & 0.43$\pm$0.13  &  -1.23$\pm$0.06   &   12.0   &       &  -1.32 \\
NGC 6304& 13:            & 0.38$\pm$0.11  &  -0.40$\pm$0.20   &          &       &  -0.59 \\
NGC 6316& 12:            & 0.37$\pm$0.16  &  -0.65$\pm$0.13   &          &       &         \\
NGC 6333& 10:            & 0.5:           &  -1.7:            &   12.0   &       &        \\
NGC 6342& 11:            & 0.41$\pm$0.14  &  -0.82$\pm$0.17   &   12.0   &   0.32&  -0.65 \\
NGC 6352& 12:            & 0.37$\pm$0.12  &  -0.48$\pm$0.19   &    9.9   &   0.44&  -0.79 \\
NGC 6356& 13:            & 0.41$\pm$0.12  &  -0.53$\pm$0.14   &          &       &  -0.50 \\
NGC 6362& 12:            & 0.5:           &  -1.1:            &   11.3   &   0.43&  -0.95 \\
NGC 6388& 12:            & 0.20$\pm$0.13  &  -0.70$\pm$0.09   &   10.6   &       &  -0.60 \\
NGC 6441& 15:            & 0.3:           &  -0.6:            &   12.7   &       &  -0.53 \\
NGC 6522& 13$\pm$1       & 0.37$\pm$0.17  &  -1.17$\pm$0.07   &   12.0   &       &        \\
NGC 6528& 15:            & 0.27$\pm$0.08  &  -0.14$\pm$0.08   &          &   0.11&  -0.04 \\
NGC 6544& 12:            & 0.21$\pm$0.22  &  -1.19$\pm$0.11   &   12.7   &       &  -1.56 \\
NGC 6553& 15$\pm$4       & 0.31$\pm$0.09  &  -0.20$\pm$0.08   &          &       &  -0.21 \\
NGC 6569& 12:            & 0.47$\pm$0.16  &  -0.87$\pm$0.17   &   10.9   &       &  -0.86 \\
NGC 6624& 12:            & 0.33$\pm$0.11  &  -0.63$\pm$0.09   &   10.6   &       &  -0.44 \\
NGC 6626& 12$\pm$3       & 0.48$\pm$0.14  &  -1.30$\pm$0.06   &   12.0   &       &  -1.45 \\
NGC 6637& 12:            & 0.44$\pm$0.13  &  -0.67$\pm$0.09   &   10.6   &       &  -0.70 \\
NGC 6638& 12$\pm$3       & 0.38$\pm$0.14  &  -0.87$\pm$0.07   &   11.5   &       &  -0.99 \\
NGC 6652& 12:            & 0.42$\pm$0.13  &  -0.92$\pm$0.13   &   10.5   &       &  -0.96 \\
NGC 6752& 10:            & 0.5:           &  -1.6:            &    8.7   &   0.33&  -1.48 \\
NGC 7078& 14$\pm$2       & 0.5:           &  -2.19$\pm$0.26   &   10.4   &   0.38&  -2.26 \\
NGC 7089& 10$\pm$3       & 0.48$\pm$0.21  &  -1.59$\pm$0.19   &          &       &  -1.63 \\
NGC 5946& 9$\pm$4        & 0.36$\pm$0.31  &  -1.52$\pm$0.29   &   12.5   &       &        \\
NGC 5986& 9$\pm$4        & 0.23$\pm$0.24  &  -1.61$\pm$0.27   &   12.0   &       &  -1.58 \\
\hline
\end{tabular}
\end{table*}

\begin{table*}
\caption{Evolutionary parameters of GCs in the Large Magellanic Cloud
calculated by us using the Lick indices published by Beasley et al. (2002),
and TMB03 models.
A colon is used when $\Delta\feh > 0.4$ dex, $\Delta$(age) $> 4$ Gyr, 
and $\Delta\afe > 0.3$ dex.
}
\label{lmc}
\begin{tabular}{lccr} \\
\hline \hline
Cluster &  age     &  \afe &   $\zh $  \\
	&   (Gyr)  & (dex) &  (dex)                 \\
\hline
NGC 1718    &     2.2$\pm$0.9 & 0.17$\pm$0.20 &-0.95$\pm$ 0.16 \\
NGC 1751    &     1.0$\pm$0.1 & 0.16$\pm$0.15 &-0.18$\pm$ 0.15 \\
NGC 1754    &     5.7$\pm$0.9 & 0.17$\pm$0.26 &-1.24$\pm$ 0.12 \\
NGC 1786    &     12.$\pm$2.6 & 0.30$\pm$0.24 &-1.50$\pm$ 0.07 \\
NGC 1801    &     0.3$\pm$0.2 & 0.22$\pm$0.24 &-0.87$\pm$ 0.34 \\
NGC 1806    &     1.5$\pm$0.2 & 0.24$\pm$0.09 &-0.49$\pm$ 0.12 \\
NGC 1830    &     1.0$\pm$1.4 & 0.37:         &-1.29$\pm$ 0.25 \\
NGC 1835    &     7.0$\pm$1.6 & 0.33$\pm$0.26 &-1.34$\pm$ 0.11 \\
NGC 1846    &     2.0$\pm$0.7 & 0.15$\pm$0.17 &-0.85$\pm$ 0.20 \\
NGC 1852    &     1.8$\pm$0.5 & 0.27$\pm$0.14 &-0.77$\pm$ 0.16 \\
NGC 1856    &     0.4$\pm$0.06& 0.13$\pm$0.13 &-0.13$\pm$ 0.12 \\
NGC 1865    &     0.5$\pm$0.08& 0.36$\pm$0.19 &-0.27$\pm$ 0.15 \\
NGC 1872    &     0.4$\pm$0.09& 0.33$\pm$0.23 &-0.64$\pm$ 0.30 \\
NGC 1878    &     0.4$\pm$0.09& 0.22$\pm$0.16 &-0.22$\pm$ 0.14 \\
NGC 1898    &     6.5$\pm$2.2 & 0.25:                 &-1.00$\pm$ 0.23 \\
NGC 1916    &     7.4$\pm$1.5 & 0.10$\pm$0.25 &-1.68$\pm$ 0.17 \\
NGC 1939    &     8.0$\pm$1.4 & 0.5:          &-1.68$\pm$ 0.16 \\
NGC 1978    &     1.8$\pm$0.3 & 0.18$\pm$0.15 &-0.29$\pm$ 0.12 \\
NGC 1987    &     1.0$\pm$0.4 & 0.33$\pm$0.22 &-0.66$\pm$ 0.24 \\
NGC 2005    &     5.8$\pm$1.2 & 0.14:                  &-1.35$\pm$ 0.31 \\
NGC 2019    &    13.5$\pm$1.3 & 0.00$\pm$0.25 &-1.58$\pm$ 0.14 \\
NGC 2107    &     0.4$\pm$0.1 & 0.13$\pm$0.14 &-0.31$\pm$ 0.16 \\
NGC 2108    &     1.0$\pm$0.4 & 0.25$\pm$0.23 &-0.64$\pm$ 0.26 \\
SL 250      &     1.0$\pm$0.4 & 0.41$\pm$0.20 &-0.93$\pm$ 0.17 \\
\hline
\end{tabular}
\end{table*}

\begin{table*}
\caption{Evolutionary parameters of GCs in the Large Magellanic Cloud
calculated by us using the Lick indices published by Beasley et al. (2002),
 and the GD05 model spectra.}
\label{lmcGD05}
\begin{tabular}{lcr} \\
\hline \hline
Cluster &  age$^{GD05}$ &  $ Z/Z_{\sun}^{GD05}$  \\
	&  (Gyr)        &  ($Z_{\sun})$          \\
\hline
NGC 1751    & 1.6$\pm$0.2  &    0.64$\pm$0.16 \\
NGC 1754    & 5.0$\pm$1.0  &    0.09$\pm$0.09 \\
NGC 1801    & 0.3$\pm$0.2  &    0.06$\pm$0.07 \\
NGC 1806    & 5.0$\pm$0.9  &    0.05$\pm$0.08 \\
NGC 1830    & 2.8$\pm$0.8  &    0.05$\pm$0.11 \\
NGC 1846    & 4.2$\pm$1.2  &    0.05$\pm$0.09 \\
NGC 1852    & 4.3$\pm$1.1  &    0.05$\pm$0.08 \\
NGC 1856    & 0.9$\pm$0.2  &    0.40$\pm$0.20 \\
NGC 1865    & 1.1$\pm$0.9  &    0.33$\pm$0.45 \\
NGC 1872    & 2.4$\pm$0.7  &    0.05$\pm$0.09 \\
NGC 1878    & 1.1$\pm$0.2  &    0.19$\pm$0.04 \\
NGC 1978    & 2.8$\pm$0.5  &    0.49$\pm$0.21 \\
NGC 1987    & 1.3$\pm$0.3  &    0.53$\pm$0.19 \\
NGC 2005    & 5.0$\pm$1.0  &    0.06$\pm$0.10 \\
NGC 2107    & 2.0$\pm$0.4  &    0.10$\pm$0.05 \\
NGC 2108    & 1.5$\pm$0.3  &    0.47$\pm$0.16 \\
SL 250      & 3.1$\pm$1.0  &    0.05$\pm$0.10 \\
\hline
\end{tabular}
\end{table*}

\end{document}